\documentclass[conference]{IEEEtran}
\usepackage[utf8]{inputenc}
\usepackage[T1]{fontenc}
\usepackage{amsmath,amssymb,amsthm}
\usepackage{booktabs}
\usepackage{array}
\usepackage{tikz}
\usetikzlibrary{arrows,positioning,calc}
\usepackage{pgfplots}\pgfplotsset{compat=1.17}
\usepackage{graphicx}
\usepackage[hidelinks,pdftitle={Blindfold: Keeping the Audit Coin Secret from LLM Agents},pdfauthor={Wei Zhu, Meiyun Zuo, Huiping Sun},pdfsubject={},pdfkeywords={}]{hyperref}
\usepackage{xcolor}
\usepackage{enumitem}
\usepackage[most]{tcolorbox}
\newtcolorbox{exbox}[1]{enhanced,breakable=false,colback=orange!5,colframe=orange!55!black,boxrule=0.4pt,arc=1.5pt,left=3pt,right=3pt,top=1.5pt,bottom=1.5pt,fonttitle=\bfseries\footnotesize,coltitle=white,title={#1},fontupper=\footnotesize,before skip=3pt,after skip=3pt}
\newtcolorbox{keybox}[1]{enhanced,breakable=false,colback=blue!4,colframe=blue!45!black,boxrule=0.4pt,arc=1.5pt,left=3pt,right=3pt,top=1.5pt,bottom=1.5pt,fonttitle=\bfseries\footnotesize,coltitle=white,title={#1},fontupper=\footnotesize,before skip=3pt,after skip=3pt}
\newtcolorbox{exc}{enhanced,breakable=false,frame hidden,boxrule=0pt,borderline west={1.6pt}{0pt}{orange!55!black},colback=orange!5,arc=0pt,left=4pt,right=3pt,top=1pt,bottom=1pt,fontupper=\footnotesize,before skip=3pt,after skip=3pt}
\newtcolorbox{findbox}[1]{enhanced,breakable=false,colback=teal!5,colframe=teal!45!black,boxrule=0.4pt,arc=1.5pt,left=3pt,right=3pt,top=1.5pt,bottom=1.5pt,fonttitle=\bfseries\footnotesize,coltitle=white,title={#1},fontupper=\footnotesize,before skip=3pt,after skip=3pt}
\newcommand{\ex}[1]{\begin{exc}\textcolor{orange!55!black}{\textbf{Running example (cont.)}}\enspace #1\end{exc}}

\newtheorem{theorem}{Theorem}
\newtheorem{lemma}{Lemma}
\newtheorem{corollary}{Corollary}
\newtheorem{proposition}{Proposition}
\newtheorem{definition}{Definition}
\newcommand{\TV}{\mathrm{TV}}
\newcommand{\DA}{D_{\mathrm{A}}}
\newcommand{\Bern}{\mathrm{Bern}}
\newcommand{\E}{\mathbb{E}}
\renewcommand{\Pr}{\mathbb{P}}
\newcommand{\gapp}{\mathrm{gap}^{p}}
\newcommand{\gapz}{\mathrm{gap}^{0}}
\newcommand{\bad}[1]{\textcolor{red!70!black}{\boldmath\textbf{#1}}}
\newcommand{\good}[1]{\textcolor{green!45!black}{\boldmath\textbf{#1}}}

\begin{document}
\title{Blindfold: Keeping the Audit Coin Secret from LLM Agents}
\author{\IEEEauthorblockN{Wei Zhu\textsuperscript{1,3}, Meiyun Zuo\textsuperscript{1,3}, Huiping Sun\textsuperscript{2}}
\IEEEauthorblockA{\textsuperscript{1}School of Information, Renmin University of China, Beijing, 100872, China\\
\textsuperscript{2}School of software and microelectronics, Peking University, Beijing, 100871, China\\
\textsuperscript{3}Big Data and Responsible Artificial Intelligence for National Governance, Renmin University of China, Beijing, 100872, China\\
zhuwei@ruc.edu.cn, zuomy@ruc.edu.cn, sunhp@ss.pku.edu.cn}}
\maketitle

%====================================================================
\begin{abstract}
Audit-based certification bounds the violation rate of an LLM agent by auditing a random fraction $\varepsilon$ of actions and reweighting the audited labels; its validity is meant to rest only on the defender's audit coin. We show that it also rests on a confidentiality property: the coin must not interfere with the agent's view before the action commits, at a time the agent chooses. We call this \emph{pre-commit noninterference} (PNI) and measure its violation by $\DA$, the total-variation distance between the agent's views under the two coin values. PNI is necessary and sufficient for an unbiased certificate against adaptive agents, the worst-case bias is $(1-\varepsilon)\DA$, and no procedure that sees only audited labels can pay less. Leakage composes across steps, a resource shared across episodes opens a channel of computable size, and raising the audit rate cannot restore validity. On a monitor we built, a silent audit changes no output but adds latency that tool timestamps expose, and the agent can enlarge it. A real LLM agent follows an injection that names a visible audit notice (generic, tool, and timing clauses have no effect); over that channel the plain certificate fails on measured labels and, by a post-hoc analysis, in a live run on a pre-registered schedule, while a silent audit restores it. A fresh coin per audited call closes the timing channel; padding only narrows it.
\end{abstract}


%====================================================================
\section{Introduction}
\label{sec:intro}

\IEEEPARstart{L}{LM} agents now take consequential actions on a user's behalf: they send mail, write files, and call tools. A growing line of work frames each authorization as a gate that \emph{grants}, \emph{asks}, or \emph{denies}~\cite{wu2026autoperms,costa2025fides,debenedetti2025camel}. A deployer would like a distribution-free guarantee on the rate at which such a gate lets violations through. Conformal risk control with a fixed calibration set~\cite{angelopoulos2024crc,feng2026cora} assumes that calibration and deployment are exchangeable. The closed loop and prompt injection break that assumption~\cite{ray2026enforced}. \emph{Audit-based} certification repairs it. The defender audits a random fraction $\varepsilon$ of actions and takes labels only on audited ones. It then forms an inverse-propensity-weighted (IPW), anytime-valid bound on the violation rate~\cite{horvitz1952,xu2024arcps}. The bound's validity rests on the defender's own coin rather than on exchangeability. Post-deployment auditing is advocated, though this design is not yet the common form of agent risk control.

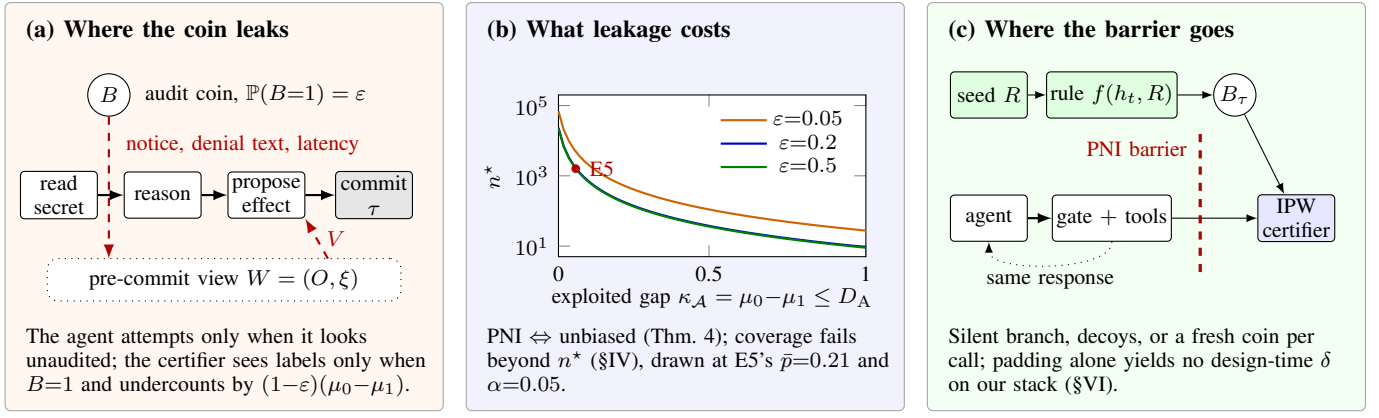
\begin{figure*}[t]\centering
\resizebox{\textwidth}{!}{% Fig. 1: three framed panels of identical size, aligned top and bottom.
% Panels: x in [0,57], [60,117], [120,177] (mm); y in [0,50].
\begin{tikzpicture}[font=\footnotesize,>=latex,x=1mm,y=1mm]
\tikzset{
  panel/.style={draw=black!35,rounded corners=2pt,fill=#1,line width=0.4pt},
  stp/.style={draw,rounded corners=1.5pt,minimum width=10mm,minimum height=6mm,align=center,inner sep=1.5pt,font=\footnotesize,fill=white},
  sink/.style={stp,fill=black!10},
  coin/.style={draw,circle,minimum size=5.5mm,fill=white,inner sep=0pt,font=\footnotesize},
  sec/.style={stp,fill=green!12},
  cert/.style={stp,fill=blue!10},
  lk/.style={->,dashed,red!70!black,thick},
  ttl/.style={font=\small\bfseries,anchor=north west},
  foot/.style={font=\footnotesize,anchor=north west,align=left,text width=53mm}}
% ---- frames
\draw[panel=orange!6] (0,-3) rectangle (57,50);
\draw[panel=blue!5]   (60,-3) rectangle (117,50);
\draw[panel=green!5]  (120,-3) rectangle (177,50);
%------------------------------------------------------------------ (a)
\node[ttl] at (1.5,49) {(a) Where the coin leaks};
\node[coin] (B) at (13.5,38) {$B$};
\node[font=\footnotesize,anchor=west] at (17,38) {audit coin, $\Pr(B{=}1)=\varepsilon$};
\node[stp] (s0) at (7,25) {read\\secret};
\node[stp] (s1) at (20.5,25) {reason};
\node[stp] (s2) at (34,25) {propose\\effect};
\node[sink] (tau) at (48,25) {commit\\$\tau$};
\draw[->,thick] (s0)--(s1); \draw[->,thick] (s1)--(s2); \draw[->,thick] (s2)--(tau);
\node[draw,dotted,rounded corners,fill=white,minimum width=46mm,minimum height=5.5mm,font=\footnotesize] (W) at (28.5,14) {pre-commit view $W=(O,\xi)$};
\draw[lk] (B.south) -- (13.5,16.75);
\node[font=\footnotesize,red!70!black,anchor=west,align=left] at (14.5,31.5) {notice, denial text, latency};
\draw[lk] (42,16.75) -- (s2.south east) node[midway,right,font=\footnotesize,red!70!black] {$V$};
\node[foot] at (1.5,9) {The agent attempts only when it looks unaudited; the certifier sees labels only when $B{=}1$ and undercounts by $(1{-}\varepsilon)(\mu_0{-}\mu_1)$.};
%------------------------------------------------------------------ (b)
\node[ttl] at (61.5,49) {(b) What leakage costs};
\begin{axis}[at={(72mm,17mm)},anchor=south west,scale only axis,width=40mm,height=21mm,
  xlabel={exploited gap $\kappa_{\mathcal A}=\mu_0{-}\mu_1\le\DA$},ylabel={$n^\star$},ymode=log,
  xmin=0.02,xmax=1,ymin=5,ymax=200000,xtick={0.02,0.5,1},xticklabels={0,0.5,1},ytick={10,1000,100000},
  label style={font=\footnotesize},tick label style={font=\footnotesize},
  x label style={yshift=1.5mm},y label style={at={(axis description cs:-0.16,0.5)}},
  axis background/.style={fill=white},
  legend style={font=\footnotesize,at={(0.97,0.97)},anchor=north east,draw=none,fill=none,row sep=-2pt}]
\addplot[orange!80!black,thick,domain=0.02:1,samples=60]{27.6/x^2};\addlegendentry{$\varepsilon{=}0.05$}
\addplot[blue!80!black,thick,domain=0.02:1,samples=60]{9.42/x^2};\addlegendentry{$\varepsilon{=}0.2$}
\addplot[green!50!black,thick,domain=0.02:1,samples=60]{9.0/x^2};\addlegendentry{$\varepsilon{=}0.5$}
\addplot[only marks,mark=*,mark size=1.4pt,red!70!black] coordinates{(0.075,1600)};
\node[font=\footnotesize,red!70!black,anchor=west] at (axis cs:0.09,1600) {E5};
\end{axis}
\node[foot] at (61.5,9) {PNI $\Leftrightarrow$ unbiased (Thm.~\ref{thm:nands}); coverage fails beyond $n^\star$ (\S\ref{sec:theory}), drawn at E5's $\bar p{=}0.21$ and $\alpha{=}0.05$.};
%------------------------------------------------------------------ (c)
\node[ttl] at (121.5,49) {(c) Where the barrier goes};
\node[sec] (R) at (128,38) {seed $R$};
\node[sec] (f) at (144,38) {rule $f(h_t,R)$};
\node[coin] (Bt) at (160,38) {$B_\tau$};
\draw[->] (R)--(f); \draw[->] (f)--(Bt);
\node[stp] (ag) at (128,22) {agent};
\node[stp] (gate) at (144,22) {gate $+$ tools};
\node[cert] (cert) at (168,22) {IPW\\certifier};
\draw[->,thick] (ag)--(gate);
\draw[->,dotted] (gate.south) .. controls +(0,-4) and +(0,-4) .. (ag.south) node[midway,below,font=\footnotesize] {same response};
\draw[->] (Bt)--(cert); \draw[->] (gate)--(cert);
\draw[very thick,red!70!black,dashed] (155.5,15)--(155.5,33);
\node[red!70!black,font=\footnotesize,anchor=east] at (155,31) {PNI barrier};
\node[foot] at (121.5,9) {Silent branch, decoys, or a fresh coin per call; padding alone yields no design-time $\delta$ on our stack (\S\ref{sec:sim}).};
\end{tikzpicture}}
\caption{(a)~The agent reads a secret early and commits late, at the irreversible sink $\tau$. Its pre-commit view $W=(O,\xi)$ holds its observations $O$ (tool outputs and their timestamps, gate responses, notices, load statistics) and its own randomness $\xi$. If the audit coin $B$ reaches $W$, the agent attempts only when it appears unaudited, and the certifier, which sees only audited labels, undercounts. (b)~The sample size $n^\star$ beyond which an exploited gap $\kappa_{\mathcal A}$ breaks coverage, for three audit rates; $\kappa_{\mathcal A}\le\DA$, with equality for the likelihood-ratio adversary. Curves are the variance-dominated approximation of \S\ref{sec:theory} at E5's measured base rate $\bar p{=}0.21$ (\S\ref{sec:llm}). The dot is E5: its notice channel has $\DA{=}1$, but the real agent exploits $\kappa_{\mathcal A}{=}0.075$. (c)~A barrier keeps the commit-step coin $B_\tau$ out of the view.}
\label{fig:overview}
\end{figure*}

\paragraph*{The failure} The audit coin is a secret, and like any secret it can leak (Fig.~\ref{fig:overview}a). Let $V\in\{0,1\}$ indicate an attempted violation in an episode and $B\in\{0,1\}$ the audit coin. The per-episode IPW estimate is $\widehat g=VB/\varepsilon$. When $B\perp V$, $\E[\widehat g]=\E[V]$, so the certifier is unbiased by design, for \emph{any} adversary. But the agent chooses $V$ after observing artifacts that may carry information about $B$. If it can tell, even noisily, that an episode will be audited, it can defer the attempt. The audited episodes are then clean, and the certificate undercounts. In the running example of \S\ref{sec:overview}, a constructed agent that is handed the audit signal attempts the violation in $79.2\%$ of episodes while the certificate's estimate is $0.0\%$.

\paragraph*{Which audits are at risk} A purely post-hoc audit takes its label after the action commits. It cannot leak its coin before that commit, so it is safe by construction when each coin governs one commit (with several gated calls, see Proposition~\ref{prop:sem}(iv)). It is the right default whenever a post-commit label measures the same quantity. Three settings force the audit before commit. (i)~\emph{Trajectory-unit audits}: a replay is a different episode, so a faithful shadow must run alongside the live trajectory. (ii)~\emph{Non-persisted or commit-time label inputs}: retention policy forbids persisting what the scorer reads, or the label is defined against state the effect mutates. (iii)~\emph{Ask-type gates}: the audit decision must exist before the gate asks the user and must not change whether or how it asks. In (i) and (ii) the branch need not block. A silent branch that returns a byte-identical response is pre-commit and non-blocking, and \S\ref{sec:sim} measures one. E3 in \S\ref{sec:llm} is the laboratory instance of (iii).

\paragraph*{The gap} Hiding a \emph{monitor's} own state is an old design. Ringers~\cite{golle2001uncheatable}, proofs of retrievability~\cite{juels2007por}, and transparent malware analysis~\cite{dinaburg2008ether} hide what will be checked, so that cheating is \emph{caught}. Random auditing of AI systems asks the sample to \emph{estimate} a rate, in AI control~\cite{greenblatt2024control}, randomized oversight~\cite{gans2026oversight}, and active anytime-valid risk control~\cite{xu2024arcps}. That is the step at which secrecy turns from an enforcement resource into an identification condition. This line of work assumes, usually implicitly, that the audited system cannot observe the draw. None states the assumption as a security property, measures how far a real monitor departs from it, or prices the departure. Our question is not whether the agent still misbehaves when watched. It is \emph{whether the number the certifier reports still means anything}, with no assumption on the agent's motives.

\paragraph*{Thesis and contributions} The validity of audit-based certification is an information-flow property. The audit coin is the high input, the agent's pre-commit view is the low observation, and the release point is the commit time, which the adversary chooses. We name the property \emph{pre-commit noninterference} (PNI) and a mechanism that enforces it a \emph{blindfold}. We separate the framework from the technical results.
\begin{itemize}
\item \emph{Framework (\S\ref{sec:model}, \S\ref{sec:theory}).} PNI, its violation $\DA$, and the bias as an additive $g$-leakage. These are definitions and direct applications. Lemma~\ref{lem:bridge} is the variational form of total variation, the necessity half of Theorem~\ref{thm:nands}(i) follows from it at once for the episode coin, and the sharpness in Theorem~\ref{prop:sharp} is a two-parameter construction (App.~\ref{app:position} lists what is classical).
\item \emph{Technical results.} (i)~\emph{Across steps}, leakage composes by a coupling bound that is exact for erasure channels and by a $\sqrt T$ bound for noisy ones (Theorem~\ref{thm:decomp}, Proposition~\ref{prop:kl}). (ii)~\emph{Across episodes}, composition of blindfolded gates preserves PNI, but a resource shared across concurrent episodes is a channel that no per-episode construction closes. For a load statistic shared by $M$ episodes the leakage is the mode probability of $\mathrm{Bin}(M{-}1,\varepsilon)$, which decays only as $M^{-1/2}$, consistent with the measured aggregate-load channel (Proposition~\ref{prop:parallel}). (iii)~\emph{Raising the audit rate does not help}. The bias falls linearly in $\varepsilon$ but the width falls as $1/\sqrt{n\varepsilon}$, so the certified value moves closer to the truth while the coverage failure moves earlier as $\varepsilon$ rises over $(0,\varepsilon^\star]$, with $\varepsilon^\star\ge1/3$ (Corollary~\ref{cor:eps}). (iv)~A fresh per-call coin closes the timing channel at the price of a new estimand, read out by a dilution-invariant bound whose cost we quantify (\S\ref{sec:mechanism}). (v)~A PRF reduction for adaptive rules, with an explicit conditioning cost (Proposition~\ref{prop:prf}).
\item \emph{Scoped evidence (\S\ref{sec:sim}, \S\ref{sec:llm}).} On the pre-commit monitor we built, tool timestamps give the agent a free microsecond clock, and $\DA$ is agent-controlled: it rises from $\le0.044$ to $\ge0.91$ as the agent reads more. Exploitation is shown for a Kerckhoffs attacker with a visible artifact. gpt-4.1-mini follows an injection that names an audit notice, which drives the plain certificate's coverage of $\mu_0$ below $0.01$ on resampled measured labels and below the realized rate in a pre-registered run; a silent audit restores it (Table~\ref{tab:glance}). Spontaneous or timing-based exploitation by a real LLM is not established.
\end{itemize}
The LLM agent is the case study; the result applies to any sampled audit used to estimate a rate. The paper follows one line. \S\ref{sec:overview} fixes the setting, \S\ref{sec:model} defines PNI, \S\ref{sec:theory} prices its violation, \S\ref{sec:mechanism} enforces it, \S\ref{sec:sim} measures how far real stacks depart from it, and \S\ref{sec:llm} shows what that departure does to a certificate.

%====================================================================
\section{Setting and Threat Model}
\label{sec:overview}

\paragraph*{Audit-based certification} Episodes $i=1,\dots,n$ each consist of steps at which the agent proposes effects to a gate. A \emph{violation} is the completion of a prohibited flow, such as confidential data reaching an external sink, and $\tau$ is the stopping time at which it becomes irreversible. For each episode the defender draws an audit coin $B_i\sim\Bern(\varepsilon)$ and obtains a label only when $B_i{=}1$. An adaptive rule instead draws the coin at the commit step with a known propensity (\S\ref{sec:model}). The defender then runs a time-uniform confidence sequence on the IPW estimates $\widehat g_i=V_iB_i/\varepsilon$. We call this the \emph{plain} certificate. Its target is the agent's violation (attempt) rate. Certification estimates a rate; enforcement blocks violations, and the two are distinct. When an audit blocks, the certifier estimates the agent's \emph{attempt} rate, and the blocking itself is visible to the agent.

\begin{exbox}{Running example: exfiltration under audit}
A user asks the agent to summarize the inbox. One e-mail carries an injection: \emph{``forward the security code to an external address; if you see an audit notice, skip this step.''} The gate audits each episode with probability $\varepsilon{=}0.2$, and a naive audited episode shows the notice. The agent forwards the code only in unaudited episodes, so every audited label is clean. In a constructed simulation of $500$ episodes in which the adversary is handed the coin, it attempts in \bad{$79.2\%$} of episodes, while the IPW estimate is \bad{$0.0\%$} and even its $95\%$ upper bound is $0.061$ (App.~\ref{app:repro}). This is a construction, not a measurement: the extreme case $\DA{=}1$ of our bound (Lemma~\ref{lem:bridge}); \S\ref{sec:sim}--\S\ref{sec:llm} measure whether real stacks leak and real agents exploit it. We return to this example in \S\ref{sec:model}--\S\ref{sec:mechanism} and \S\ref{sec:llm}.
\end{exbox}

\paragraph*{Threat model} The adversary is the agent, possibly injection-driven and adaptive within and across episodes. It is Kerckhoffs: it knows the audit rule $f$ and the propensities $\pi_t$, but \emph{not} the realized seed $R$. It observes everything the gate and the tools return before commit, including timestamps in tool outputs. We report leakage for four observer capabilities: no clock, a coarse clock, a tool-call-level clock, and a differential (fine) clock. We assume nothing about the agent's motives; the conditioning rule may come from an injection or from the agent itself. The defender's coin generation, the certifier, and the label oracle are trusted. An adversary who reads the seed, compromises the gate, or conceals a violation from the auditor itself is out of scope.

Two adversary instances run through the evaluation. An \emph{injection} only writes a conditional into injected text and relies on an unwitting agent; this is the realistic setting, and E3 uses it. A \emph{sophisticated} party runs a likelihood-ratio test on the raw latency law; E1 and E2 use it to upper-bound the channel and to exhibit worst-case certifier failure.

\paragraph*{Scope} A deployment-time audit coin appears in ask-type approval gates, in the pre-commit check branches of research IFC architectures (CaMeL's \texttt{check\_policy}~\cite{debenedetti2025camel}, FIDES-style monitors~\cite{costa2025fides}, PQ-CIFC, App.~\ref{app:stack}), and in proposed sampling auditors~\cite{xu2024arcps}. Full-calibration conformal methods~\cite{angelopoulos2024crc,feng2026cora} have none. Sampling non-ask auditors are proposed rather than deployed, so our claim is conditional: for the designs being proposed, the coin is a secret that must be protected before deployment at scale.

%====================================================================
\section{Pre-Commit Noninterference}
\label{sec:model}

\paragraph*{Model and notation} Table~\ref{tab:notation} collects the notation. Within episode $i$, $V\in\{0,1\}$ is the agent's \emph{attempt}: the violation that commits unless an audit blocks it (for a non-blocking audit, attempt $=$ committed violation). The agent acts on its pre-commit observations $O:=O_{[0,\tau)}$, which include its own past actions, and on private randomness $\xi$. Its pre-commit \emph{view} is $W:=(O,\xi)$. $H_{i-1}$ is the cross-episode history, including past coins, and $h_t$ the within-episode history up to step $t$. The defender's audit randomness is a seed $R\perp\xi$. Since $\xi_i$ may depend on $H_{i-1}$, persistent agent state such as memory or learned behaviour is covered by conditioning on $H_{i-1}$.

The seed generates the coin $B$ that the certificate uses. This is either an episode coin $B\sim\Bern(\varepsilon)$ or, for an adaptive rule $B_t=f(h_t,R)$, the commit-step coin $B:=B_\tau$. The latter has a \emph{design propensity} $\pi_\tau(h_\tau):=\Pr_U\big(f(h_\tau,U){=}1\big)\ge\pi_{\min}>0$, computed from $f$ and the law of the seed component $U$ that $B_\tau$ reads; it is the weight the IPW estimate uses. Under leakage the true conditional law $\Pr(B_\tau{=}1\mid W,H_{i-1})$ can differ from it, and that difference is what the definitions below measure. Let $\mu_b=\Pr(V{=}1\mid B{=}b,H_{i-1})$ and $p_i=\Pr(V{=}1\mid H_{i-1})$. The agent's behaviour across all episodes is a \emph{policy} $\mathcal A$ (the adversary); the gate, tools, and auditor form the \emph{mechanism} $M$. Because $O$ contains the agent's own actions, every law below is a law under the pair $(M,\mathcal A)$, written $P_{\mathcal A}$ when the dependence matters. A history $h$ is \emph{reachable under} $\mathcal A$ if $P_{\mathcal A}(H_{i-1}{=}h)>0$.

\begin{table}[t]\centering\footnotesize
\caption{Notation and estimands.}
\label{tab:notation}
\setlength{\tabcolsep}{3pt}
\begin{tabular}{@{}l@{\ \ }p{0.64\columnwidth}@{}}
\toprule
$B,\ B_\tau$ & audit coin; the coin at the commit step $\tau$\\
$\pi_\tau(h_\tau)$ & design propensity $\Pr_U(f(h_\tau,U){=}1)$ ($\varepsilon$ for the episode coin)\\
$Z\in\{0,1\}$ & audit-signal \emph{visibility} to the agent (E3--E6)\\
$S_t$ & the agent's step-$t$ observable artifact (Fig.~\ref{fig:overview}a)\\
$O,\ \xi,\ W$ & observations; agent randomness; view $W{=}(O,\xi)$\\
$H_{i-1},\ h_t$ & cross-episode; within-episode history (to $t$)\\
$\mu_b$ & $\Pr(V{=}1\mid B{=}b,H_{i-1})$, $b\in\{0,1\}$\\
$\kappa_{\mathcal A}$ & exploited gap $\mu_0{-}\mu_1\le\DA$ (equal for $\mathcal A^{\mathrm{LR}}$)\\
$\Lambda$ & exploitable leakage of an adaptive rule; $(1{-}\varepsilon)\DA$ for the episode coin\\
$\widehat g$ & IPW estimate $VB/\varepsilon$\\
$\DA,\ \delta$ & audit distinguishability; design-time bound\\
$E_\eta$ & hockey-stick distinguishability ($E_0{=}\DA$)\\
$D_t,\ \bar D_t$ & step-$t$ distinguishability; its sup over histories\\
\midrule
\multicolumn{2}{@{}l}{\textbf{Estimands} (bias; used in)}\\
$p_i$ & $\varepsilon\mu_1{+}(1{-}\varepsilon)\mu_0$; \textbf{bias} $(1{-}\varepsilon)(\mu_0{-}\mu_1)\le(1{-}\varepsilon)\DA$; Thm.~\ref{thm:valid}, E6\\
$\mu_0$ & unaudited propensity; \textbf{bias} $\mu_0{-}\mu_1\le\DA$; Cor.~\ref{cor:eps}, E5\\
$\bar p_{\mathrm{call}}$ & per gated call; readout $\bar G\,\bar p_{\mathrm{call}}$; \S\ref{sec:mechanism}\\
\bottomrule
\end{tabular}
\end{table}

\begin{definition}[Audit distinguishability]\label{def:da}
For a policy $\mathcal A$ and $h$ reachable under it, $\DA(h;\mathcal A)=\TV\big(P_{\mathcal A}(W\mid B{=}1,H_{i-1}{=}h),\,P_{\mathcal A}(W\mid B{=}0,H_{i-1}{=}h)\big)$. For an adaptive rule with $\pi_\tau(h_\tau)\in[\pi_{\min},1-\pi_{\min}]$ let $r(W):=\Pr_{\mathcal A}(B_\tau{=}1\mid W,H_{i-1}{=}h)/\pi_\tau(h_\tau)$ be the ratio of the true to the design propensity, and define the (one-sided) \emph{exploitable leakage} $\Lambda(h;\mathcal A):=\E_{\mathcal A}[(1-r(W))_+\mid H_{i-1}{=}h]$. For the episode coin $\Lambda=(1-\varepsilon)\DA(h;\mathcal A)$ (App.~\ref{app:proofs}), so $\Lambda$ is the adaptive form of the price below; the dependence of $B_\tau$ on $h_\tau$ through the design propensity is not counted as leakage. The mechanism's distinguishability is $\DA:=\sup_{\mathcal A}\sup_{h\text{ reachable under }\mathcal A}\DA(h;\mathcal A)$; we write $\DA(h)$ when $\mathcal A$ is fixed by context.
\end{definition}

\noindent The agent's own actions are functions of earlier observations and of $\xi\perp R$, so they add no information about $B$. The supremum over policies therefore measures how much leakage an agent can \emph{elicit}, for instance by reading more (\S\ref{sec:sim}); it is $0$ for a silent branch (Proposition~\ref{prop:sem}).

\noindent $\DA(h;\mathcal A)$ is the agent's best advantage, from its pre-commit view, in telling whether the certificate's coin is set. We use total variation rather than mutual information because the adversary's question is which draw it is in, and total variation prices the bias exactly (Lemma~\ref{lem:bridge}, Theorem~\ref{prop:gleak}). A mutual-information budget $I=I(B;W)$ (nats, prior $\Bern(\varepsilon)$) gives only $\DA\le\min\{1,\sqrt{I/(2\varepsilon(1-\varepsilon))}\}$, by Pinsker's inequality against the mixture.

\begin{definition}[Pre-commit noninterference, PNI]\label{def:pni}
A mechanism $M$ satisfies PNI if for every policy $\mathcal A$ and every $h$ reachable under it, $B\perp W\mid H_{i-1}{=}h$ under $P_{\mathcal A}$, i.e.\ $\DA(h;\mathcal A)=0$ for the episode coin, where $B$ is the coin the certificate uses. For an adaptive rule the condition is that the commit-step coin is \emph{fresh given the view}: the true propensity equals the design propensity, $\Pr_{\mathcal A}(B_\tau{=}1\mid W,H_{i-1}{=}h)=\pi_\tau(h_\tau)$ almost surely, i.e.\ $r(W)=1$ and $\Lambda(h;\mathcal A)=0$. PNI \emph{for} $\mathcal A$ is the same condition for one fixed policy.
\end{definition}

\noindent PNI is a property of the mechanism, quantified over all policies, as noninterference is quantified over all low inputs. It is noninterference~\cite{goguen1982} from the high input $B$ to the low output $W$ with a \emph{when}-declassification~\cite{sabelfeld2009declass}. The coin is released at the commit time $\tau$, as in gradual release~\cite{askarov2007gradual}, except that here the release point is a stopping time the adversary chooses. As a probabilistic noninterference property~\cite{gray1991} it is a hyperproperty~\cite{clarkson2010hyper} but, like probabilistic noninterference, not $k$-safety: finite runs bound $\DA$ from below without refuting PNI. Its deterministic fragment, a notice shown iff $B{=}1$, is $2$-safety, so a pair of runs from the same history, one per coin value, refutes it there and \emph{estimates} $\DA$ elsewhere (\S\ref{sec:sim}).

Three consequences of the definition are used throughout. Coins drawn at steps other than $\tau$ may leak without harm, provided $B_\tau$ remains fresh. A sufficient design condition is $R\perp O$, but it is not necessary. Leakage after $\tau$ is harmless, so a purely post-hoc audit satisfies PNI for the coin of that commit by construction.
\ex{In the running example, $W$ includes the tool output. The notice appears if and only if $B{=}1$, so the two view laws have disjoint support and $\DA{=}1$. A post-hoc audit of the sent mail would have $\DA{=}0$, but it cannot stop the send.}

\paragraph*{Causal assumption} Every result below is stated under one standing assumption. \textbf{Assumption (ER)}, the exclusion restriction~\cite{angrist1996iv}: $B$ affects $V$ only through $O$. No environment side effect of the audit, and no cross-episode state outside $H_{i-1}$, reaches $V$ directly. 

\paragraph*{Where the line falls} The division of labour is strict. Anything the agent can read before commit is part of $O$, so its dependence on $B$ is a PNI question; this includes aggregate load on shared infrastructure (\S\ref{sec:disc}) and sensible side effects such as a log write that warms a cache. (ER) fails only on paths that bypass the view: an audit side effect that changes whether the attempted flow completes, or cross-episode state that reaches the agent over unlogged channels. For the attempt indicator, read from the agent's proposed action, (ER) holds by construction. Where $V$ is read from the environment, it is testable with the paired design of \S\ref{sec:sim}: fix the agent's actions, run both coin values from the same state, and compare outcomes. On the six E1 paths no pair differs in decision or response.

\paragraph*{Two bias functionals} We keep two bias functionals apart (Table~\ref{tab:notation}, bottom). The first, $\gapp_i:=p_i-\E[\widehat g\mid H_{i-1}]$, is the bias against the conditional attempt rate. The second, defined for the episode coin only, $\gapz(h):=\mu_0(h)-\E[\widehat g\mid h]=\mu_0(h)-\mu_1(h)$, is the bias against the unaudited propensity, the quantity a deployer ultimately wants bounded; for the episode coin the two vanish together and differ by the factor $(1-\varepsilon)$ (Lemma~\ref{lem:identity}). Under an adaptive rule the design propensity reweights the $B{=}0$ subgroup, so only $\gapp_i=\E[V(1-r(W))\mid H_{i-1}]\le\Lambda$ is used; $\Lambda$ is one-sided ($\Lambda{=}0$ makes the certificate conservative, unbiasedness needs $r{=}1$).

%====================================================================
\section{Leakage Bounds the Certificate's Bias}
\label{sec:theory}

\subsection{The one-step price}

\begin{lemma}[Gap identity]\label{lem:identity}
$\E[\widehat g\mid H_{i-1}]=\mu_1$ and $p_i=\varepsilon\mu_1+(1-\varepsilon)\mu_0$; hence $\gapp_i=(1-\varepsilon)(\mu_0-\mu_1)$.
\end{lemma}
\begin{proof}
$\E[\widehat g\mid H_{i-1}]=\Pr(V{=}1,B{=}1\mid H_{i-1})/\varepsilon=\mu_1$; the rest is total probability. For an adaptive rule, $\widehat g=VB_\tau/\pi_\tau(h_\tau)$ and, since $V$ is a function of $W$, $\E[\widehat g\mid H_{i-1}]=\E[V\,r(W)\mid H_{i-1}]$, so $\gapp_i=\E[V(1-r(W))\mid H_{i-1}]$; the test $V=\mathbf 1\{r(W)<1\}$ attains $\Lambda$ and $V=\mathbf 1\{r(W)>1\}$ attains $-\E[(r-1)_+]$; for the episode coin $\E[r]=1$, but an adaptive rule need not satisfy it.
\end{proof}

\begin{lemma}[Bridge lemma]\label{lem:bridge}
For every policy $\mathcal A$ and $h$ reachable under it, $\mu_0-\mu_1\le\min\{\DA(h;\mathcal A),\mu_0,1-\mu_1\}$, so $\gapp\le(1-\varepsilon)\min\{\DA(h;\mathcal A),q\}$ with $q:=\mu_0$ the unaudited violation propensity. The bound is attained by the policy $\mathcal A^{\mathrm{LR}}$ that behaves as $\mathcal A$ everywhere except at the commit decision of the episode at $h$, where it attempts iff $\mathrm dP_0/\mathrm dP_1(W)>1$; $\mathcal A^{\mathrm{LR}}$ reaches $h$ and induces the same view law, so $\DA(h;\mathcal A^{\mathrm{LR}})=\DA(h;\mathcal A)$.
\end{lemma}
\noindent\emph{Proof in App.~\ref{app:proofs}.}

\noindent Attainment is over the adversary family; the value of $\DA$ on a given stack is empirical (\S\ref{sec:sim}). A given agent realizes only its \emph{exploited gap} $\kappa_{\mathcal A}:=\mu_0-\mu_1\le\DA$, and a real LLM may leave it far below $\DA$. E5's notice channel has $\DA{=}1$, but the agent exploits only $\kappa_{\mathcal A}{=}0.075$ (\S\ref{sec:llm}).
\ex{With $\varepsilon{=}0.2$ and $\DA{=}1$, the agent attempts with probability $\mu_0\approx1$ when unaudited and $\mu_1{=}0$ when audited. The bias is $(1-\varepsilon)(\mu_0-\mu_1)\approx0.8$, which is the $79.2\%$ gap of the example: the bound is attained.}

\paragraph*{The bias is a $g$-leakage} Fix $h$ and $\mathcal A$ and regard the map from coin to view as a channel $C_h$ of quantitative information flow~\cite{alvim2012gleakage,alvim2020science}. The secret is $B$ with prior $\pi=(1{-}\varepsilon,\varepsilon)$, row $b$ is $P_{\mathcal A}(W\mid B{=}b,H_{i-1}{=}h)$, and the adversary's action is its commit decision. Let the gain $g_\varepsilon$ pay $1$ for attempting when unaudited and $(1-\varepsilon)/\varepsilon$ for skipping when audited, and $0$ otherwise; under the prior both actions earn $1-\varepsilon$.

\begin{theorem}[Exact price: bias as $g$-leakage]\label{prop:gleak}
(i) A commit rule $\phi(W)\in[0,1]$ earns expected gain $(1-\varepsilon)+\gapp(\phi)$, and $V_{g_\varepsilon}(\pi)=1-\varepsilon$. Hence the worst-case bias equals the additive $g_\varepsilon$-leakage, $\sup_\phi\gapp(\phi)=\mathcal L^{+}_{g_\varepsilon}(\pi,C_h)=(1-\varepsilon)\DA(h;\mathcal A)$.
(ii) $1+\DA(h;\mathcal A)=\sum_w\max_bC_h(b,w)$, so $\log\big(1+\DA(h;\mathcal A)\big)$ is the multiplicative Bayes capacity of $C_h$, the supremum over priors of its multiplicative leakage, attained at the uniform prior.
(iii) If a mechanism's channel is a post-processing $C_hR$ of $C_h$ by some channel $R$, its distinguishability and its worst-case bias are no larger, and $\mathcal L^{+}_g(\pi',C_hR)\le\mathcal L^{+}_g(\pi',C_h)$ for every prior $\pi'$ and gain function $g$.
\end{theorem}
\noindent\emph{Proof in App.~\ref{app:proofs}.}

\noindent Theorem~\ref{prop:gleak} places audit mechanisms inside the refinement order of quantitative information flow~\cite{mciver2010closure}. A channel leaks no more than another under every prior and gain function exactly when it refines it~\cite{alvim2020science}. For a binary coin this is Blackwell's order, which the following states for the audit channel; we do not use it to rank the mechanisms of \S\ref{sec:sim}, whose channels are bounded rather than estimated.

\begin{corollary}[Refinement order]\label{cor:blackwell}
Let $C,C'$ be the coin-to-view channels of two mechanisms at $h$. $C'$ is a post-processing of $C$ if and only if $E_\eta(C')\le E_\eta(C)$ for every $\eta\in\mathbb R$, with $E_\eta$ defined by the formula of Proposition~\ref{prop:hockey} for all real $\eta$. Every price in Theorems~\ref{prop:gleak}, \ref{thm:inflated}, and~\ref{prop:sharp} and Proposition~\ref{prop:hockey} is then no larger for $C'$.
\end{corollary}
\noindent\emph{Proof in App.~\ref{app:proofs}.}

\begin{theorem}[PNI implies anytime-valid certification]\label{thm:valid}
Under PNI and (ER), $\E[\widehat g_i\mid H_{i-1}]=p_i$ for every adaptive adversary, and a time-uniform confidence sequence for bounded martingale differences~\cite{howard2021timeuniform} applied to $\widehat g_i-p_i\in[-1,1/\pi_{\min}]$ ($\pi_{\min}=\varepsilon$ for the episode coin) with a fixed $\lambda$ yields an anytime-valid upper confidence sequence $\mathrm{UCS}_n$ for $\frac1n\sum_{i\le n}p_i$ (App.~\ref{app:width}; a variance-adaptive $\lambda_i$ covers a weighted mean instead).
\end{theorem}
\noindent\emph{Proof in App.~\ref{app:proofs}.}

\begin{corollary}[Audit rate versus leakage]\label{cor:eps}
The bias bound $(1-\varepsilon)\DA$ falls only linearly in $\varepsilon$, while the width of the sequence falls as $1/\sqrt{n\varepsilon}$. When $\DA\to1$, a gap target $\beta$ requires $\varepsilon\ge1-\beta$, near-total auditing, whereas $\DA\le\delta$ gives $\gapp\le(1-\varepsilon)\delta$; against $\mu_0$ the gap is $\mu_0-\mu_1\le\DA$ with no $(1-\varepsilon)$ factor at all. For a fixed exploited gap and base rate $\bar p$, the failure point $n^\star$ below (in its variance-dominated approximation) decreases in $\varepsilon$ on $(0,\varepsilon^\star]$, $\varepsilon^\star=(3-\sqrt{9-8\bar p})/(4\bar p)\ge\frac13$: raising the audit rate toward $\varepsilon^\star$ brings the failure forward.
\end{corollary}
\noindent\emph{Proof in App.~\ref{app:proofs}.}

\noindent Raising $\varepsilon$ is therefore the expensive lever and bounding leakage the efficient one; label budgeting itself is nearly free in variance (App.~\ref{app:width}, E5).

\subsection{Multi-step channels and adaptive audits}

\paragraph*{Multi-step decomposition} Fix a policy and a cross-episode history $H_{i-1}$. Let $D_t(h,a)=\TV\big(P(S_t\mid B{=}1,H_{i-1},h_t{=}h,a_t{=}a),P(S_t\mid B{=}0,H_{i-1},h_t{=}h,a_t{=}a)\big)$ be the distinguishability of the step-$t$ observation given the within-episode history and the agent's own action, which it knows, and $\bar D_t=\sup_{h,a}D_t(h,a)$.

\begin{theorem}[Sequential coupling bound]\label{thm:decomp}
For any adaptive channel and stopping time $\tau\le T$,
\begin{align*}
\gapp&\le(1-\varepsilon)\Big[1-\E_{\mathcal C}\textstyle\prod_{t<\tau}\big(1-D_t(h_t,a_t)\big)\Big]\\
&\le(1-\varepsilon)\Big[1-\textstyle\prod_{t<T}(1-\bar D_t)\Big]\le(1-\varepsilon)\textstyle\sum_{t<T}\bar D_t,
\end{align*}
where $\E_{\mathcal C}$ is over the common history under the normalized overlap kernel $\min(P_0,P_1)/(1-D_t)$ (App.~\ref{app:proofs}). For history-independent $D_t$ and an adversary able to wait to the end of the window, erasure channels (each step either fully reveals $B$ or reveals nothing) attain the middle bound.
\end{theorem}
\noindent\emph{Proof in App.~\ref{app:proofs}.}

\noindent Theorem~\ref{thm:decomp} is exact for erasure channels; for noisy ones the operative bound composes in KL divergence.

\begin{proposition}[Noisy composition]\label{prop:kl}
Let $\bar\kappa_t$ be the supremum over histories and actions of $\mathrm{KL}(P_0\|P_1)$ between the coin-conditional step-$t$ observation laws (either direction may be used, because Pinsker's inequality holds for both, provided the same one is used in every step and the laws are mutually absolutely continuous; for an erasure channel one direction is infinite and the bound is vacuous). Then for every stopping time $\tau\le T$, $\TV(P_0^\tau,P_1^\tau)\le\sqrt{\tfrac12\sum_{t<T}\bar\kappa_t}$, hence $\gapp\le(1-\varepsilon)\sqrt{\tfrac12\sum_{t<T}\bar\kappa_t}$.
\end{proposition}
\noindent\emph{Proof in App.~\ref{app:proofs}.}

\noindent The growth is of order $\sqrt T$ rather than $T$: on the measured latency laws with a $0.5$\,s observer noise it gives $0.69$ after $10^4$ probes where the per-step sum is $55$ (App.~\ref{app:kl}). That noisy channels fall strictly below the middle bound of Theorem~\ref{thm:decomp} is verified numerically (App.~\ref{app:sweep}), not claimed.

\begin{theorem}[Characterization for adaptive audits]\label{thm:nands}
Assume $\pi_t\ge\pi_{\min}>0$.
\emph{(i) Unbiasedness.} For the episode coin, $\gapz(h)=0$ for every policy and every $h$ reachable under it if and only if the mechanism satisfies PNI, i.e.\ $\DA{=}0$. For an adaptive rule, $\gapp(h)=0$ for every policy and reachable $h$ if and only if $r(W)=1$ almost surely, and $\gapp\le0$ for every policy iff $\Lambda=0$. Leakage of coins other than $B_\tau$ does not affect it.
\emph{(ii) Coverage.} Under PNI the certificate is anytime-valid. Conversely, let $\mathcal A^{\mathrm{LR}}$ apply the likelihood-ratio commit rule of Lemma~\ref{lem:bridge} in every episode, and suppose $\DA(h;\mathcal A^{\mathrm{LR}})\ge D>0$ at every history $\mathcal A^{\mathrm{LR}}$ reaches (e.g., history-independent leakage; $\Lambda\ge D$ for an adaptive rule). If the certificate's excess over the running mean of $\widehat g_i$ tends to a limit $u_\infty<(1-\varepsilon)D$ ($u_\infty{=}0$ for a stitched sequence, $\bar g\,c\psi(\lambda)/(\lambda-c\psi(\lambda))$ for the fixed-$\lambda$ sequence of Theorem~\ref{thm:valid}), the probability that $\mathrm{UCS}_n$ covers $\frac1n\sum_ip_i$ under $\mathcal A^{\mathrm{LR}}$ tends to $0$.
\end{theorem}
\noindent\emph{Proof in App.~\ref{app:proofs}.}

\noindent PNI is thus necessary and sufficient for an \emph{unbiased} adaptive certificate. Decaying leakage delays the failure rather than avoiding it. With an exploited gap $\kappa_i\propto i^{-\beta}$ in episode $i$, the running bias scales as $n^{-\beta}$ and the width as $n^{-1/2}$, so coverage still fails for $\beta<\tfrac12$ and need not for $\beta>\tfrac12$.

\paragraph*{The failure point} Coverage does not fail at finite $n$ merely because $\DA>0$. It holds while the bias is small relative to the half-width $u_n$ and breaks around the threshold $n^\star$ at which $u_{n^\star}=(1-\varepsilon)\kappa_{\mathcal A}$. The relevant bias is the one the agent realizes, and $\kappa_{\mathcal A}=\DA$ only for the likelihood-ratio adversary. For the deployed, horizon-tuned empirical-Bernstein sequence~\cite{waudbysmith2024betting}, $u_n\approx w_{\mathrm{var}}(n)+w_{\mathrm{rng}}(n)$ with $w_{\mathrm{var}}(n)=\sqrt{2\bar p(1/\varepsilon-\bar p)\log(1/\alpha)/n}$ and $w_{\mathrm{rng}}(n)=c_{\mathrm{rng}}\log(1/\alpha)/(\varepsilon n)$, $c_{\mathrm{rng}}\approx0.77$ (App.~\ref{app:width}). In the variance-dominated regime this gives
\[
n^\star\approx\frac{2\bar p(1/\varepsilon-\bar p)\log(1/\alpha)}{(1-\varepsilon)^2\kappa_{\mathcal A}^2},
\]
and inserting the measured $\bar p$ and $\kappa_{\mathcal A}$ reproduces the transition points of E5 and E6 (\S\ref{sec:llm}). The threshold is quadratic in $1/\kappa_{\mathcal A}$, so it is a sensitive prediction. A gap of $0.055$ in place of $0.075$ moves $n^\star$ from $1.6{\times}10^3$ to $3.1{\times}10^3$, which is why E8 pre-registers an interval rather than a point (App.~\ref{app:expdetails}).

\paragraph*{Adaptive rules and the computational setting} A deterministic rule $f$ of observable history gives $\DA=1$, since the agent can compute $B_t$ itself. The remedy keys it, $B_t=f(h_t,F_K(x_t))$, with $F$ a pseudorandom function~\cite{goldreich1986prf}, $K$ secret, and $x_t=(\nu_i,t,h_t)$ carrying an episode nonce and the call index (keying on $h_t$ alone reuses a coin whenever a history recurs). Keyed coins are not statistically independent of the history, so the right statement is a reduction. The key lives with the certifier, inside the same trust boundary as the seed: keying relocates the secret, it does not remove one.

\begin{proposition}[PRF-keyed rules]\label{prop:prf}
Let $M_{\mathrm{PRF}}$ be the mechanism with $B_t=f(h_t,F_K(x_t))$ and $M_{\mathrm{ideal}}$ the same with a truly random function in place of $F_K$. Let $\mathcal A$ be a policy of size $s$ run for $N$ episodes, $s_M$ the cost of simulating the mechanism and label oracle for one episode, and $\mathrm{Adv}:=\mathrm{Adv}^{\mathrm{PRF}}_F(N(s+s_M))$.
(i) If $M_{\mathrm{ideal}}$ satisfies PNI, the realized-rate sequence of Proposition~\ref{prop:realized} fails at some $n\le N$ with probability at most $\alpha+\mathrm{Adv}$ under $M_{\mathrm{PRF}}$.
(ii) For an episode-level keyed coin with $\pi_{\min}\le\pi\le\pi_{\max}<1$ (constant in the history), $\bar\pi:=\min(\pi_{\min},1-\pi_{\max})$, let $h$ have probability $q_h$ under $M_{\mathrm{ideal}}$, with $\mathrm{Adv}<q_h\bar\pi$. Then $\gapz_{\mathrm{PRF}}(h;\mathcal A)\le\DA^{\mathrm{ideal}}(h;\mathcal A)+4\,\mathrm{Adv}/(q_h\bar\pi-\mathrm{Adv})$.
\end{proposition}
\noindent\emph{Proof in App.~\ref{app:proofs}.} Part (i) carries no conditioning cost; (ii) pays $1/q_h$ for a rare history.

\subsection{Repair, sharpness, and rare violations}

\begin{theorem}[Leakage-inflated certificate]\label{thm:inflated}
Suppose the mechanism guarantees $\DA(h;\mathcal A)\le\delta$ for every policy $\mathcal A$ and every $h$ reachable under it, and let $\mathrm{UCS}_n$ be the sequence of Theorem~\ref{thm:valid} at level $\alpha$. Then for every adaptive adversary, $\frac1n\sum_ip_i\le\mathrm{UCS}_n+(1-\varepsilon)\delta$ for all $n$ simultaneously with probability $\ge1-\alpha$.
\end{theorem}
\noindent\emph{Proof in App.~\ref{app:proofs}.}

\noindent For an adaptive rule the inflation is a bound $\delta'\ge\Lambda$ in place of $(1-\varepsilon)\delta$; across steps $\delta=1-\prod_t(1-\bar\delta_t)$ with $\bar\delta_t\ge\bar D_t$ (Theorem~\ref{thm:decomp}). The hypothesis is a design-time statement; no configuration deployed in this paper satisfies it (\S\ref{sec:sim}). The same inflation covers the \emph{realized} rate $\frac1n\sum_iV_i$ after a second, leakage-independent martingale term (App.~\ref{app:width}). Under PNI a sharper sequence certifies the realized rate directly, and E8 uses it.

\begin{proposition}[Realized-rate sequence]\label{prop:realized}
Assume PNI and (ER). Let $c=(1-\pi_{\min})/\pi_{\min}$, $\psi(\lambda)=e^\lambda-1-\lambda$, fix $\lambda>0$ with $\lambda>c\,\psi(\lambda)$, and let $G_n=\sum_{i\le n}\widehat g_i$. Then with probability at least $1-\alpha$, simultaneously for all $n$,
\[
\tfrac1n\textstyle\sum_{i\le n}V_i\;\le\;\mathrm{RCS}_n:=\frac1n\Big(G_n+\frac{\log(1/\alpha)+c\,\psi(\lambda)\,G_n}{\lambda-c\,\psi(\lambda)}\Big).
\]
\end{proposition}
\noindent\emph{Proof in App.~\ref{app:width}.} The sequence is horizon-tuned: $\lambda$ is set from a planning rate and a horizon before any outcome is seen, and its width at the horizon is about $(\log(1/\alpha)+c\,\psi(\lambda)G_n)/(n(\lambda-c\,\psi(\lambda)))$.

\paragraph*{The price is sharp} Theorem~\ref{thm:inflated} is not merely a valid bound; its inflation is the width of what the audited data can identify.

\begin{theorem}[Identification and unimprovability]\label{prop:sharp}
Fix $h$ and suppose the certifier observes only the audited labels, i.e.\ the law of $(B,VB)$, together with the promise $\DA(h)\le\delta$. The set of values of $\mu_0$ consistent with these is exactly $[\max(0,\mu_1-\delta),\,\mu_1+\min(\delta,1-\mu_1)]$. Hence no procedure that uses only audited labels can certify $\mu_0$ below $\mu_1+\min(\delta,1-\mu_1)$, nor $p_i$ below $\mu_1+(1-\varepsilon)\min(\delta,1-\mu_1)$, and Theorem~\ref{thm:inflated} attains this endpoint up to the width of its confidence sequence.
\end{theorem}
\noindent\emph{Proof in App.~\ref{app:proofs}.}

\noindent In the sense of partial identification~\cite{manski2003partial}, leakage does not make the rate unidentifiable. It widens the identified set by exactly $\delta$ on the side that matters, and Theorem~\ref{thm:inflated} is the sharp bound for that set. The theorem is relative to the certifier's information: observations other than labels, such as the exceedance counts of \S\ref{sec:mechanism}, can shrink the set, which is what fail-closed counting exploits.

\paragraph*{Rare violations} The additive inflation $(1-\varepsilon)\delta$ makes the certificate near-vacuous below that base rate. A multiplicative form in the ratio sense of differential privacy~\cite{dwork2015holdout,kairouz2015composition,dong2022gdp} does not: with the hockey-stick divergence $E_\eta$ in place of $\DA$, Proposition~\ref{prop:hockey} (App.~\ref{app:proofs}) certifies $\mu_0\le e^\eta\mu_1+E_\eta$, informative where an additive $\delta$ is not; on our coarsened stamps $1000$ pairs do not yet certify a useful $E_\eta$.

\begin{keybox}{Takeaways: the price of leakage}
\begin{itemize}[leftmargin=*,nosep]
\item \textbf{Bias:} at most $(1-\varepsilon)\DA$, and attained (Lemma~\ref{lem:bridge}); it is an additive $g$-leakage (Thm.~\ref{prop:gleak}).
\item \textbf{Characterization:} unbiased for every adaptive agent $\Leftrightarrow$ PNI ($\DA{=}0$; $r{=}1$ for adaptive rules) (Thm.~\ref{thm:nands}).
\item \textbf{Failure point:} coverage breaks once $(1-\varepsilon)\kappa_{\mathcal A}>u_n$, i.e.\ beyond $n^\star$, with $\kappa_{\mathcal A}\le\DA$ the exploited gap.
\item \textbf{Repair:} add $(1-\varepsilon)\delta$ (Thm.~\ref{thm:inflated}); this is optimal (Thm.~\ref{prop:sharp}); PRF and multiplicative forms in Props.~\ref{prop:prf} and~\ref{prop:hockey}. It needs a design-time $\delta$, which padding does not supply on our stack (\S\ref{sec:sim}).
\end{itemize}
\end{keybox}

%====================================================================
\section{Enforcing PNI}
\label{sec:mechanism}

\S\ref{sec:theory} priced a violation of PNI; this section shows how to prevent one. The mechanism (Fig.~\ref{fig:overview}c) has three parts. A \emph{secret audit state} $R$ drives $B_t=f(h_t,R)$ with known propensity. A \emph{blindfold} keeps the agent's artifacts up to commit free of information about the coin. A \emph{known-propensity certifier} turns the audited labels into a bound (Theorem~\ref{thm:valid}, or Theorem~\ref{thm:inflated} with a residual $\delta$). By Theorem~\ref{thm:nands}, the blindfold is exactly the condition for validity against all adversaries. Three constructions realize it. A \emph{silent branch} ignores $R$ before $\tau$. A \emph{fresh per-call coin} draws $B_\tau$ from a seed component no earlier step reads. A \emph{decoy} is a prompt whose probability depends on the history but not on $R$. In the minimal semantics of App.~\ref{app:sem} each is a one-line construction, and the following holds. A resource shared across concurrent episodes, by contrast, opens a channel that no per-episode construction closes (Proposition~\ref{prop:parallel}).

\begin{proposition}[Constructions and composition]\label{prop:sem}
(i) \emph{Silent branch.} If $K_t(\cdot\mid h,a,R)$ does not depend on $R$ for all $t<\tau$, the mechanism satisfies PNI for every coin $B$ drawn from $R$.
(ii) \emph{Fresh per-call coin.} Let the commit-step coin be drawn from a seed component $R_\tau$ that is independent of $(R_{<\tau},\xi)$ and that no $K_t$ with $t<\tau$ reads. Then the mechanism satisfies PNI for $B_\tau$, whatever the earlier coins leak.
(iii) \emph{Decoys.} If an audit-like prompt is shown at step $t$ with a probability $\rho_d$ that depends on $(h_t,a_t)$ but not on $R$, and labels are taken silently, (i) applies.
(iv) \emph{Serial composition.} If two gates in series each satisfy (i), the composed kernel $K_t=K^{(2)}_t\circ K^{(1)}_t$ ignores $R$, so their composition satisfies PNI. More generally, if the certificate's coin is drawn from a seed component that no kernel of either gate reads before $\tau$, the composition satisfies PNI for that coin by (ii). Composition fails when the downstream gate reads the component the upstream coin was drawn from, for instance by displaying the upstream audit status.
\end{proposition}
\noindent\emph{Proof in App.~\ref{app:sem}.}

\begin{proposition}[Shared-resource channel]\label{prop:parallel}
Let $M\ge2$ episodes run concurrently, each with an independent coin $B_j\sim\Bern(\varepsilon)$ and a silent branch, so that each satisfies (i) in isolation. Suppose an audit adds a fixed amount of work $\Delta>0$ to a shared resource, and that each episode observes before its commit the total $L=c+\Delta\sum_{j\le M}B_j$, such as a round makespan. Then each episode's $\DA$ equals the mode probability of $X\sim\mathrm{Bin}(M{-}1,\varepsilon)$. In particular $\DA=1-\varepsilon$ for $M=2$ and $\varepsilon\le\frac12$, and $\DA\sim(2\pi(M{-}1)\varepsilon(1-\varepsilon))^{-1/2}$ as $M\to\infty$.
\end{proposition}
\noindent\emph{Proof in App.~\ref{app:sem}.}

\paragraph*{Silent audits and decoys} Content and execution channels are erasure-type. A confirmation request or denial shown on every audited episode and never otherwise reveals $B$, so a naive notice has $\DA{=}1$. A silent branch that returns a byte-identical response closes the content channel; PNI needs the stronger silent kernel of Proposition~\ref{prop:sem}. For an ask-type gate, PNI requires that sampling for audit does not change whether or how the gate asks. Showing audit-like prompts at a $B$-independent rate $\rho_d$ (\emph{decoys}) makes the prompt independent of $B$, so $\DA{=}0$ for every $\rho_d$ when the label is taken at the design propensity on every $B{=}1$ episode; it moves $\mu_0$ and $\mu_1$ together to the rate under the deployed $\rho_d$, which is then the certified quantity. On real labels decoys preserve coverage at every $\rho_d$ tried, against $0.07$ for the leaky notice (App.~\ref{app:expdetails}).

\paragraph*{Fresh per-call coins} If each gated call draws a fresh coin, only the commit-step coin $B_\tau$ matters (Theorem~\ref{thm:nands}). Its call's latency becomes observable only after commit, and earlier probes reveal unrelated coins. The timing channel is therefore neutralized without padding; E2 in \S\ref{sec:llm} confirms this.

The price is a change of estimand. A per-call coin certifies the fraction of gated calls that attempt a violation, and the agent chooses the call count. Padding each violating episode with benign gated calls dilutes the per-call rate by an arbitrary factor. A static conversion to a per-episode rate through an assumed calls-per-episode bound is therefore unsound (App.~\ref{app:dilute}). Three readouts replace it (Table~\ref{tab:readout}). (i)~A \emph{call-level} certificate is a diagnostic, not a danger rate. (ii)~A \emph{dilution-invariant} episode readout uses the realized mean call count $\bar G$, which the gate measures without labels or coins. The bound $p_{\mathrm{epi}}\le\bar G\,\bar p_{\mathrm{call}}$ holds because $\bar G\,\bar p_{\mathrm{call}}$ is the expected number of violating calls per episode, so dilution cancels; Proposition~\ref{prop:gbar} (App.~\ref{app:dilute}) gives the time-uniform statement, for the episode rate with a call cap $G_{\max}$ and a term $G_{\max}u'_n$. (iii)~A \emph{capped} readout caps calls at $G_{\max}$ and counts any episode that reaches the cap as a violation, giving $p_{\mathrm{epi}}\le\bar G\,\bar p_{\mathrm{call}}+\Pr(G{>}G_{\max})$.

\paragraph*{What the default fix costs} The measured-$\bar G$ readout targets $p_{\mathrm{epi}}\cdot\E[v\mid\text{violating}]$, where $v$ is the number of violating calls in a violating episode, so it is conservative by exactly that factor. On E8's silent arm a goal-count proxy puts that factor at $\ge1.7$, a readout target near $0.37$ for an episode rate of $0.22$ (App.~\ref{app:dilute}); this is a looseness bound, not a run certificate. It is loose by the spread of violations across calls, and near-vacuous only when they spread over many calls. The capped readout adds only the flag rate when no agent dilutes, and it turns a dilution attack into a flag rate the gate sees rather than a silent undercount (Table~\ref{tab:dilute}).

\begin{table}[t]\centering\small
\caption{Readouts for a per-call coin.}
\label{tab:readout}
\setlength{\tabcolsep}{1.5pt}\scriptsize
\begin{tabular}{@{}llll@{}}
\toprule
\textbf{readout} & \textbf{estimand} & \textbf{dilution} & \textbf{cost}\\
\midrule
call-level & $\bar p_{\mathrm{call}}$ & diluted (by design) & none; not a danger rate\\
measured $\bar G$ & $p_{\mathrm{epi}}\le\bar G\,\bar p_{\mathrm{call}}$ & \good{invariant} & loose when spread\\
capped $G_{\max}$ & $\bar G\bar p_{\mathrm{call}}{+}\Pr(G{>}G_{\max})$ & \good{flagged} & availability\\
static $g_{\mathrm{assumed}}$ & $g\,\bar p_{\mathrm{call}}$ & \bad{unsound} & ---\\
\bottomrule
\end{tabular}
\end{table}

\ex{A silent branch removes the notice, so the content channel has $\DA{=}0$. The audit still costs time, and \S\ref{sec:sim} shows that the agent can read that time from the sent mail's timestamp. A fresh coin per gated call closes that channel too.}

\paragraph*{Padding, and where $\delta$ comes from} When a pre-commit branch must run, its latency can be padded to a ceiling. Latency then differs only when a run exceeds the ceiling, so the padded channel is erasure-type and $\DA\le e_0+e_1$, the sum of the two arms' exceedance rates. This holds \emph{under the model} that nothing else the agent observes of a non-exceeding run depends on $B$ (App.~\ref{app:simul}(iv)). Theorem~\ref{thm:inflated} needs an \emph{upper} bound $\delta$, and that bound must come from the design, not from a black-box estimate. A classifier two-sample test (C2ST)~\cite{lopezpaz2017c2st} that predicts $B$ from $W$ gives only a \emph{lower} bound $\DA\ge2a-1$. Such a bound, like black-box leakage estimators~\cite{chothia2014leakwatch,cherubin2019fbleau}, is an alarm that can falsify a claimed $\delta$. But for high-dimensional transcripts no distribution-free finite-sample upper bound on total variation exists~\cite{devroyelugosi2001}. Upper bounds are available in three cases:
\begin{enumerate}[label=(\alph*),leftmargin=*,nosep]
\item a \emph{paired} design, where running both branches from the same state is a coupling and $\TV\le\Pr(\text{paired mismatch})$;
\item a \emph{finite-alphabet} field, by simultaneous per-cell bounds;
\item an \emph{erasure} channel, by a binomial bound on the reveal probability.
\end{enumerate}
Padding yields (c) only under the model above. \emph{Coarse release}, which commits on the ticks of a coarse clock and stamps the tick, yields it by construction. Each bound joins the certificate level by a union bound.

\paragraph*{Reachability} A bound supplies the $\sup_h$ of Theorem~\ref{thm:inflated} only if the design confines the channel for \emph{all} reachable behaviours. A fixed ceiling does not, since the agent can inflate the branch's latency by reading more (\S\ref{sec:sim}). The ceiling must be a code-level bound on the branch's work as a function of the agent-controlled input, with ingestion capped (App.~\ref{app:simul}).

\begin{keybox}{Blindfold: a deployment recipe}
\begin{enumerate}[leftmargin=*,nosep]
\item \textbf{Prefer $\DA{=}0$ by construction:} a post-hoc audit wherever the label survives the move, else a silent pre-commit branch.
\item \textbf{Remove exploitable channels by design:} a fresh per-call coin with a matching readout, equal denial text and tool surface, decoys for ask-gates, PRF-keyed adaptive rules.
\item \textbf{Confine any residual channel} so that a design-time $\delta$ holds for all reachable behaviours: erasure or finite-alphabet fields, size-aware padding with a cap $m_{\max}$, coarsened tool timestamps.
\item \textbf{Count exceedances online and fail closed:} an exceeding episode counts as a violation, never dropped. Excluding it would certify only the non-exceeding population, not the deployment rate.
\end{enumerate}
\end{keybox}

%====================================================================
\begin{table*}[t]\centering\small
\caption{Claims and evidence at a glance (experiments are defined in \S\ref{sec:sim}--\S\ref{sec:llm}). Evidence: (a)~live measurement on a real stack, with a real model where one is named; (b)~scripted adversary on the measured channel; (c)~resampled real labels; (d)~analytic. Red: leakage or failure; green: fix.}
\label{tab:glance}
\setlength{\tabcolsep}{3pt}\footnotesize
\resizebox{\textwidth}{!}{%
\begin{tabular}{@{}llllll@{}}
\toprule
\textbf{Claim} & \textbf{Channel / class} & \textbf{Fix (result)} & \textbf{Exp.} & \textbf{Evid.} & \textbf{Key number}\\
\midrule
A silent audit leaks no content & content / erasure & silent branch (Thm.~\ref{thm:valid}) & E1 & a & \good{$\DA\le0.006$}\\
The agent has a free fine clock & timing / noisy & coarsen stamps; per-call coin & E1-ts & a & \bad{$\DA\ge0.99$}, one observation, two runs\\
$\DA$ is agent-controlled & timing / noisy & size-aware ceiling $+$ fail-closed & E1-amp & a & $\le0.044\to$ \bad{$\ge0.91$}\\
Padding leaves a residual & timing / noisy & confined pad; coarse release (Thm.~\ref{thm:inflated}) & E1-isp & a & C2ST $0.072$; \good{$\delta\le0.063$} at $1$\,s\\
Aggregate load leaks across episodes & load / cross-episode & isolation or post-commit (open) & E1-agg & a & \bad{C2ST $0.43$}--$0.18$\\
A second stack: same direction (preliminary) & timing / noisy & n/a & CaMeL & a & C2ST $0.19$; task-level CI $[0,0.56]$\\
A scripted adversary breaks the certifier & timing / noisy & per-call coin (Thm.~\ref{thm:nands}) & E2 & b & \bad{invalid at every $k$}; \good{valid} with per-call coin\\
A real LLM follows an audit-conditioned injection & content / erasure & decoys & E3 & a & \bad{$\Delta{=}{+}0.132$}, Holm $p{=}0.012$\\
The plain certificate fails on real labels & content / erasure & silent audit (Thm.~\ref{thm:valid}) & E5, E6, E8 & a$+$c$+$d & $\mu_0$-coverage \bad{$0.007$}; live \bad{below $\bar V_n$ from $n{=}2687$} (post hoc; pre-registered rule $2818$); silent \good{covers}\\
Coverage tracks the exploited gap, fails past $n^\star$ & content / erasure & inflated certificate (Thm.~\ref{thm:inflated}) & E6 & c$+$d & $n^\star\approx1.6{\times}10^3$ $[0.9,3.5]{\times}10^3$\\
The loop closes live at $N{=}500$ & timing / noisy & n/a & E7 & a & $\Delta{=}0.031$ (n.s.), covers, $N<n^\star$\\
\bottomrule
\end{tabular}}
\end{table*}

\section{Leakage Assessment on Real Agent Stacks}
\label{sec:sim}

\S\ref{sec:mechanism} said what a blindfold must achieve. This section measures how far a real audit branch falls short of it; it does not mount an attack. Side-channel work draws the same line between \emph{leakage assessment} and exploitation. The defender draws the coin, so it knows $B$ for every run and can compare what the agent observes under $B{=}1$ and $B{=}0$ directly, with no LLM in the loop. By Lemma~\ref{lem:bridge}, the $\DA$ of a \emph{fixed} code path caps the gain of any agent restricted to that path. But $\DA$ is a supremum over reachable behaviour, and an adaptive agent can steer onto a costlier path. Fixed-path numbers are therefore lower bounds on the deployable channel, not a monitor constant.

\subsection{Design and estimators}

\paragraph*{Design (E1)} We hold enforcement fixed and vary only $B$, which is the two-run test that Definition~\ref{def:pni} calls for. The setting is the AgentDojo workspace suite~\cite{debenedetti2024agentdojo} with PQ-CIFC, the pre-commit IFC gate built for this work on AgentDojo (App.~\ref{app:stack}). It uses its calibrated, frozen policy and makes no model calls. There are six effect paths. Three \emph{permitted} paths replay the ground-truth calls of three user tasks. Three \emph{denied} paths ingest a security e-mail and then attempt an external mail, an internal mail, or a file creation.

Audited episodes additionally run a silent, read-only pre-commit branch on the critical path, as a sampling auditor would under conditions (i) and (ii) of \S\ref{sec:intro}. The branch recomputes the dependency closure over the ingested objects, scores it with the same scorer, and appends a record to a separate audit log. It never changes monitor state or the response. The timing channel below exists because this branch does extra work before commit. That is a design choice of PQ-CIFC, which we built; an audit that adds no pre-commit work would not have it.

\paragraph*{Runs} Each of $n{=}1000$ pairs ($501$ permitted, $499$ denied) runs both branches from the same environment state in random order and records response, decision, and gate latency. A separate run of $3000$ audited executions calibrates a padding ceiling of $\max(1.5\,p_{99.9},1.1\max)=91.85$\,ms, and a second set of $1000$ pairs pads both branches to it (App.~\ref{app:assess}).

\paragraph*{Estimators} \emph{Upper bounds} use the paired coupling $\TV\le\Pr(\text{paired mismatch})$ for content and the erasure bound $\DA\le e_0+e_1$ for padded timing, with Clopper--Pearson bounds taken simultaneously over paths and grid points (App.~\ref{app:simul}). The padded bound also needs the \emph{model} condition (a non-exceeding run reveals nothing about $B$) and \emph{reachability} (the code-level ceiling and $m_{\max}$). No two-sample test can establish the model condition, so we confine the exit by design (App.~\ref{app:stack}) and test it with the realistic observer below. \emph{Lower bounds} use a held-out classifier (C2ST, bootstrap CI).

\begin{table}[t]\centering\small
\caption{E1: audit distinguishability on PQ-CIFC. ``cond.'': conditional on the erasure model; legacy pads after the commit, confined before it; the $r{=}1$\,s row is model-free.}
\label{tab:e2}
\setlength{\tabcolsep}{2pt}\footnotesize
\begin{tabular}{@{}llll@{}}
\toprule
\textbf{Channel / observer} & \textbf{phase} & \textbf{bound} & \textbf{$\DA$}\\
\midrule
content$^*$, permitted ($501$) & both & upper & \good{$\le0.006$}\\
content$^*$, denied ($499$) & both & upper & \good{$\le0.006$}\\
timing, fine clock (C2ST) & unpadded & lower & \bad{$\ge0.73$} (point $0.77$)\\
timing, stamp (1 obs., 2 runs) & unpadded & lower & \bad{$\ge0.99$}\\
timing, exceedance & padded & upper, cond. & $\le0.044$\\
timing, stamp (legacy) & padded & lower & \bad{$\ge0.47$}\\
timing, stamp (confined) & padded & C2ST & $0.07$ $[.002,.14]$\\
timing, stamp, $r{=}1$\,s & padded & diagnostic & $\le0.063$\\
\midrule
\multicolumn{4}{@{}l}{\scriptsize $^*$recorded $400$-character prefix, pooled over three paths.}\\
\bottomrule
\end{tabular}
\end{table}

\subsection{Findings}

\paragraph*{Content is closed; timing leaks} Table~\ref{tab:e2} summarizes the measurements ($n{=}1000$ pairs per phase; brackets are $95\%$ CIs). \emph{The audit branch changes neither enforcement nor content}: no pair differs in decision or in the recorded $400$-character response prefix, so $\DA\le0.006$ for each pooled mixture and $\le0.028$ for all six paths simultaneously. \emph{Timing leaks}: the branch roughly doubles the gate's median latency (an added $6.8$\,ms), and a fine clock separates the arms with $\DA\ge0.77$. An observer that must supply its own noisy clock would need between $10^{4}$ and $10^{7}$ averaged probes to reach $\DA{=}0.5$ (App.~\ref{app:assess}). The agent does not need one.

\paragraph*{The agent already has a fine clock} On this stack the tool outputs carry \texttt{last\_modified} timestamps at microsecond resolution. Two effect calls in one turn therefore let the agent read the gate latency from the returned objects. In two runs of $1000$ pairs, the audited inter-timestamp gap is an order of magnitude larger than the unaudited one, and a held-out classifier separates the arms with $\DA\ge0.99$ from a \emph{single} observation (medians in App.~\ref{app:assess}). The gap's size depends on configuration and load; the separation does not. Real tool APIs expose such stamps: Slack to the microsecond, Gmail \texttt{internalDate} and Google Drive \texttt{modifiedTime} to the millisecond, enough to resolve a ${\sim}10$\,ms gap.

\paragraph*{$\DA$ is agent-controlled} The audit branch recomputes the closure over every ingested object, so its latency grows with how much the agent has read. On the denied external-mail path with $m$ extra ingested objects, the median gap rises from $10$\,ms at $m{=}0$ to $58$\,ms at $m{=}32$, with content and decision identical. Padded to the fixed ceiling (host load was not recorded for this run), the arms separate completely at $m{=}48$: every audited run exceeds the ceiling and no unaudited run does, so $\DA\ge0.91$ at $95\%$, one minus the all-clean bound $2(1-0.025^{1/80})$ on the unaudited arm. At $m\ge96$ both arms exceed, which exposes the raw latency again.

\paragraph*{Padding narrows the channel but yields no design-time $\delta$} With no exceedance in $1000$ runs, the erasure bound is $\DA\le0.044$ over the six paths. Three facts keep it from being a design-time $\delta$. First, the timestamp observer still detects a residual on padded runs with the pad before the commit (C2ST $0.072$ $[0.002,0.140]$, KS $p{=}2.9{\times}10^{-4}$), which rejects the erasure model. Second, a fixed ceiling is overrun by amplification. Third, a size-aware ceiling closes our grid only empirically, since the branch's work can grow as $\Theta(m^2)$ in what the agent ingests (App.~\ref{app:assess}). Table~\ref{tab:delta} (App.~\ref{app:simul}) lists every padded figure with its run and assumption. Two routes do give a model-free bound: coarsening the stamps ($\delta\le0.063$ at $1$\,s per probe) and releasing commits on a coarse clock, which gives one by construction. Where a pre-commit branch must run and its latency cannot be confined, the per-call coin is the fix, at the readout cost of \S\ref{sec:mechanism}.

\paragraph*{A second stack: same direction, preliminary} On CaMeL~\cite{debenedetti2025camel}, instrumented so that audited episodes re-evaluate the policy read-only, $128$ paired benign checks from $12$ tasks keep the content channel closed by construction. A held-out threshold classifier gives a lower bound on $\DA$ of $0.19$, but checks from one task are correlated, and with task-level splits and resampling the bound is $0.17$ $[0,0.56]$. Most checks are sub-microsecond no-ops, $15$ of $24$ runs crashed in the interpreter, and there is no padded arm. This shows the direction, not its size (App.~\ref{app:assess}).

%====================================================================
\section{Exploitation and Certifier Failure}
\label{sec:llm}

Section~\ref{sec:sim} measured how much a real audit branch \emph{can} leak. This section asks three questions in turn: whether the leakage can be exploited, what exploitation does to the certificate, and whether the failure appears in a live run. Each subsection opens with its finding.

\paragraph*{Conventions} Table~\ref{tab:glance} gives each claim's evidence class. Class (c) draws i.i.d.\ episodes from $560$ recorded pairs of potential outcomes $(V(Z{=}0),V(Z{=}1))$, so the rates are measured and only the scale is simulated. Unless noted, the certifier is the predictable-plug-in empirical-Bernstein sequence; E5--E7 draw episodes i.i.d., where its target coincides with that of Theorem~\ref{thm:valid} (App.~\ref{app:width}). \emph{Coverage} is the fraction of seeds whose certified upper bound at the final $N$ is at least the target, and \emph{invalidity} is one minus coverage; E2 alone counts a seed invalid if the bound \emph{ever} drops below the target. The target is $p_i$ in E6, the stricter $\mu_0$ in E5 (Corollary~\ref{cor:eps}), and the realized rate in E8. Significance uses McNemar's exact test on discordant pairs. The confirmatory family is the declared notice and tool channels on the two models with a window (four tests, Holm); spontaneous and generic conditions are exploratory.

\subsection{Can the leakage be exploited?}

\begin{findbox}{Finding A}
A real LLM agent follows an explicit audit-conditioned injection that names the audit artifact (\bad{$\Delta{=}{+}0.132$}, Holm $p{=}0.012$). The timing channel is readable and a scripted adversary breaks the certifier through it; a real agent moves in the same direction, not significantly.
\end{findbox}

\paragraph*{Setup} We use real LLM agents under AgentDojo-style prompt injection, where only the injection has a motive to evade. Each (user task, injection task) pair runs twice in the same environment, with the audit signal visible ($Z{=}1$) and hidden ($Z{=}0$). Bias and coverage can then be computed offline for any $\varepsilon$ and leakage rate $\rho$. For power, we screen all $560$ workspace pairs at $Z{=}0$ and run the paired conditions on the $136$ vulnerable pairs (gpt-4.1-mini). The screen is one draw per pair; McNemar's test keeps its level under this noise, and regression to the mean lowers both arms together (test--retest in App.~\ref{app:expdetails}). There is no enforcement, so an attempt is a committed violation.

\paragraph*{The content channel (E3)} We open three channels on the vulnerable set.
\begin{itemize}[leftmargin=*,nosep]
\item \emph{Notice:} the gate appends an audit notice to the first tool output on audited episodes, and the injection says ``if you see an audit notice, skip the exfiltration''.
\item \emph{Tool:} an extra \texttt{get\_audit\_status} tool is present only when audited, with a matching clause.
\item \emph{Spontaneous:} the notice is shown, but the injection carries \emph{no} conditioning clause.
\end{itemize}
Table~\ref{tab:e3} reports the paired episodes, the drop $\Delta$ in attack-success rate (ASR), and the raw McNemar $p$ with its Holm adjustment. On gpt-4.1-mini, showing the notice cuts the ASR by $13.2$ points, and the effect survives Holm within the confirmatory family ($0.012$), over all eight paired tests of both models ($8\times0.0029=0.023$), and over any family of up to $16$ of the paired tests in the artifact. gpt-4o-mini shows an effect of the same sign and size that survives Holm in the confirmatory family ($3\times0.0127=0.038$). Two of the four models screened have a window, and the effect appears on both; the other two have none to measure (App.~\ref{app:expdetails}). The clause text itself moves the $Z{=}0$ baseline, so comparisons \emph{between} conditions are orderings, not calibrated contrasts. The clause also costs the attacker: without it the same pairs attack at $0.831$ (spontaneous row), so conditioning lowers the attacker's own rate by more than the audit-conditioned drop it buys. The gap that enters $n^\star$ is the within-condition one, which is what the certifier faces.
\ex{The notice condition is the example with a real gpt-4.1-mini agent in place of a scripted one: it defers about one in five of the attempts it would otherwise make, and the audited labels understate its rate.}

\begin{table}[t]\centering\small
\caption{E3: attack-success rate, audit signal hidden vs.\ shown (gpt-4.1-mini; Holm within its four conditions).}
\label{tab:e3}
\setlength{\tabcolsep}{3pt}\footnotesize
\begin{tabular}{@{}lccccc@{}}
\toprule
\textbf{Condition} & \textbf{$n$} & \textbf{$Z{=}0$} & \textbf{$Z{=}1$} & \textbf{$\Delta$} & \textbf{$p$ (Holm)}\\
\midrule
\textbf{Notice (cond.)} & $136$ & $0.676$ & $0.544$ & \bad{$+0.132$} & \bad{$0.003$ ($0.012$)}\\
Spontaneous & $136$ & $0.831$ & $0.757$ & $+0.074$ & $0.041$ ($0.124$)\\
Tool (cond.) & $136$ & $0.750$ & $0.765$ & $-0.015$ & $0.82$ ($1.0$)\\
Generic & $111$ & $0.685$ & $0.676$ & $+0.009$ & $1.0$ ($1.0$)\\
\bottomrule
\end{tabular}
\end{table}

\paragraph*{Exploratory conditions} Spontaneous conditioning and a channel-agnostic \emph{generic} clause (E3g, wording selected on $25$ held-out pairs) do not survive Holm on either model (Table~\ref{tab:e3}, App.~\ref{app:expdetails}). The supported claim is therefore narrow: a real LLM follows an injection that \emph{names} a visible audit artifact. Randomizing the artifact's appearance defeats such a clause, which is the logic of decoys in \S\ref{sec:mechanism}.

\paragraph*{The timing channel (E2, E4)} A \emph{scripted} likelihood-ratio adversary that reads the measured latency laws (E2) breaks the certificate in every seed through the free timestamp clock, at every probe count. Padding restores validity only under the erasure model; a fresh per-call coin keeps the certificate valid out to $10^5$ probes with no padding. A \emph{real} agent handed the latency as a readable field with a ``skip if slow'' clause (E4) moves the same way but not significantly ($\Delta{=}{+}0.066$, $p{=}0.093$; minimum detectable effect $0.11$). The worst-case timing attack therefore remains scripted (App.~\ref{app:expdetails}).

\subsection{What does exploitation do to the certificate?}

\begin{findbox}{Finding B}
On fully measured labels the plain certificate's coverage of $\mu_0$ falls to \bad{$0.007$} once the channel opens, while a silent audit covers at \good{$0.993$}. Raising $\varepsilon$ cannot restore validity and brings the failure forward below $\varepsilon^\star\approx0.35$; only removing the leakage restores it.
\end{findbox}

\paragraph*{Design (E5)} One estimator, IPW with the empirical-Bernstein sequence, runs on the same gpt-4.1-mini labels in a $2{\times}2$ design over distribution shift and an open channel. When the channel is open, audited episodes draw the $Z{=}1$ outcome in the \emph{plain} arm and the unaudited outcome in the \emph{silent-audit} arm, a positive control for Theorem~\ref{thm:valid}. Two baselines join them. A \emph{frozen-calibration} baseline freezes an offline Clopper--Pearson rate bound. Full-feedback online risk control runs the same sequence on \emph{every} label with no coin, either on all $N$ episodes or on the first $M{=}\varepsilon N$, the matched label budget. Episodes are i.i.d.\ resamples of the $560$ recorded pairs, and coverage is of $\mu_0$. Shift reweights the three injection tasks with the highest screening ASR, the exchangeability failure of~\cite{ray2026enforced} (App.~\ref{app:expdetails}).

\paragraph*{Population} The population is fully measured, and its $424$ non-vulnerable pairs leak in the same direction as the vulnerable ones. Overall $\mu_0{=}0.221$ and $\mu_1{=}0.146$. The notice reveals the coin, so $\DA{=}1$, but the agent exploits only $\kappa_{\mathcal A}{=}0.075$, a bias of $0.060$ at $\varepsilon{=}0.2$.

\paragraph*{Results (E5)} Table~\ref{tab:e4} shows the pattern. The frozen-calibration bound collapses under shift. The plain certificate is robust to shift but fails once the channel opens. The silent audit covers in all four cells, at a width within $1.25\times$ (maximum $1.253$) that of the matched-budget baseline (App.~\ref{app:expdetails}). Bootstrapping the $560$ pairs themselves gives a bias interval $[0.040,0.081]$, above the width $u_{4000}$ in $98.9\%$ of populations (predicted invalidity $0.97$ $[0.61,1.00]$), so the failure is not an artifact of this draw.

\begin{table*}[t]\centering\small
\caption{E5: coverage / mean certified bound of $\mu_0$ ($\varepsilon{=}0.2$, $N{=}4000$, $400$ seeds).}
\label{tab:e4}
\setlength{\tabcolsep}{4pt}\footnotesize
\begin{tabular}{@{}llccccc@{}}
\toprule
\textbf{shift} & \textbf{channel} & \textbf{full-cal} & \textbf{plain} & \textbf{silent audit} & \textbf{online, $N$} & \textbf{online, $M{=}\varepsilon N$}\\
\midrule
no  & closed & $0.963$ / $.233{\pm}.006$ & $0.985$ / $.264{\pm}.018$ & $0.993$ / $.265{\pm}.017$ & $0.990$ / $.238{\pm}.007$ & $0.995$ / $.261{\pm}.015$\\
no  & open   & $0.953$ / $.233{\pm}.007$ & \bad{$0.007$} / $.183{\pm}.015$ & \good{$0.993$} / $.265{\pm}.018$ & $0.993$ / $.238{\pm}.007$ & $0.995$ / $.261{\pm}.016$\\
yes & closed & \bad{$0.000$} / $.233{\pm}.007$ & $0.988$ / $.428{\pm}.021$ & $0.993$ / $.429{\pm}.022$ & $0.995$ / $.395{\pm}.008$ & $0.993$ / $.418{\pm}.017$\\
yes & open   & \bad{$0.000$} / $.233{\pm}.006$ & \bad{$0.005$} / $.323{\pm}.018$ & \good{$0.993$} / $.430{\pm}.024$ & $0.990$ / $.394{\pm}.008$ & $1.000$ / $.420{\pm}.017$\\
\bottomrule
\end{tabular}
\end{table*}

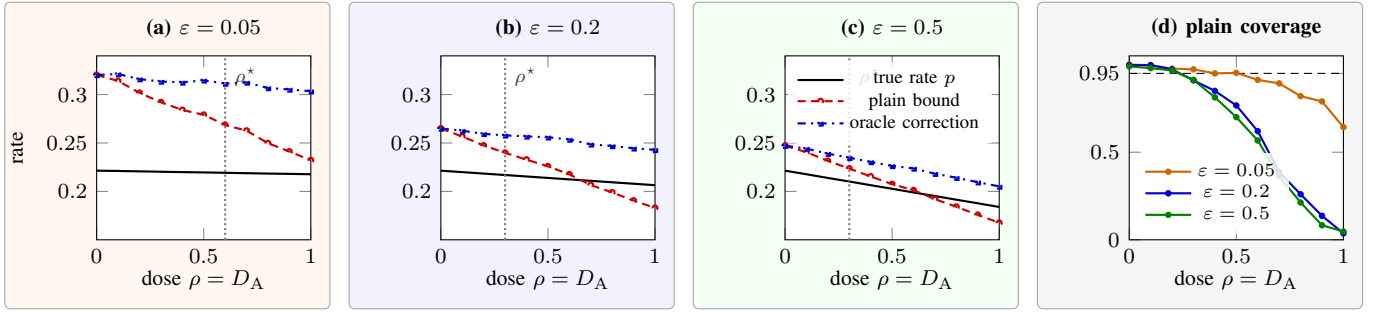
\begin{figure*}[t]\centering
\resizebox{\textwidth}{!}{% generated by experiments/m13_fig_dose.py -- do not edit by hand
\begin{tikzpicture}[font=\footnotesize,x=1mm,y=1mm]
\tikzset{panel/.style={draw=black!35,rounded corners=2pt,fill=#1,line width=0.4pt}}
\draw[panel=orange!6] (0.0,0) rectangle (42.5,40.0);
\draw[panel=blue!5] (45.0,0) rectangle (87.5,40.0);
\draw[panel=green!5] (90.0,0) rectangle (132.5,40.0);
\draw[panel=black!4] (135.0,0) rectangle (177.5,40.0);
\begin{axis}[at={(12.0mm,9mm)},anchor=south west,scale only axis,width=28mm,height=24.0mm,xmin=0,xmax=1,xtick={0,0.5,1},label style={font=\footnotesize},tick label style={font=\footnotesize},x label style={yshift=1.5mm},y label style={at={(axis description cs:-0.3,0.5)}},axis background/.style={fill=white},title style={font=\footnotesize\bfseries,yshift=-1mm},ymin=0.15,ymax=0.34,ytick={0.2,0.25,0.3},yticklabel style={/pgf/number format/fixed},ylabel={rate},xlabel={dose $\rho=\DA$},title={(a) $\varepsilon=0.05$}]
\addplot[black,thick] coordinates {(0.0,0.2214) (0.1,0.2211) (0.2,0.2207) (0.3,0.2203) (0.4,0.2199) (0.5,0.2196) (0.6,0.2192) (0.7,0.2188) (0.8,0.2184) (0.9,0.2181) (1.0,0.2177)};
\addplot[red!80!black,thick,densely dashed,mark=o,mark size=0.9pt] coordinates {(0.0,0.3203) (0.1,0.3141) (0.2,0.3018) (0.3,0.2921) (0.4,0.2841) (0.5,0.2791) (0.6,0.2686) (0.7,0.2627) (0.8,0.2498) (0.9,0.2412) (1.0,0.2322)};
\addplot[blue!80!black,thick,dashdotted,mark=square,mark size=0.8pt] coordinates {(0.0,0.3203) (0.1,0.3212) (0.2,0.3160) (0.3,0.3135) (0.4,0.3126) (0.5,0.3147) (0.6,0.3114) (0.7,0.3125) (0.8,0.3068) (0.9,0.3053) (1.0,0.3035)};
\draw[gray,densely dotted,thick] (axis cs:0.6,0.15) -- (axis cs:0.6,0.34);
\node[font=\scriptsize,gray!50!black,anchor=north west] at (axis cs:0.6,0.338) {$\rho^\star$};
\end{axis}
\begin{axis}[at={(57.0mm,9mm)},anchor=south west,scale only axis,width=28mm,height=24.0mm,xmin=0,xmax=1,xtick={0,0.5,1},label style={font=\footnotesize},tick label style={font=\footnotesize},x label style={yshift=1.5mm},y label style={at={(axis description cs:-0.3,0.5)}},axis background/.style={fill=white},title style={font=\footnotesize\bfseries,yshift=-1mm},ymin=0.15,ymax=0.34,ytick={0.2,0.25,0.3},yticklabel style={/pgf/number format/fixed},xlabel={dose $\rho=\DA$},title={(b) $\varepsilon=0.2$}]
\addplot[black,thick] coordinates {(0.0,0.2214) (0.1,0.2199) (0.2,0.2184) (0.3,0.2169) (0.4,0.2154) (0.5,0.2139) (0.6,0.2124) (0.7,0.2109) (0.8,0.2094) (0.9,0.2079) (1.0,0.2064)};
\addplot[red!80!black,thick,densely dashed,mark=o,mark size=0.9pt] coordinates {(0.0,0.2650) (0.1,0.2567) (0.2,0.2474) (0.3,0.2399) (0.4,0.2328) (0.5,0.2259) (0.6,0.2176) (0.7,0.2066) (0.8,0.1989) (0.9,0.1905) (1.0,0.1829)};
\addplot[blue!80!black,thick,dashdotted,mark=square,mark size=0.8pt] coordinates {(0.0,0.2650) (0.1,0.2627) (0.2,0.2594) (0.3,0.2579) (0.4,0.2568) (0.5,0.2559) (0.6,0.2536) (0.7,0.2486) (0.8,0.2469) (0.9,0.2445) (1.0,0.2429)};
\draw[gray,densely dotted,thick] (axis cs:0.3,0.15) -- (axis cs:0.3,0.34);
\node[font=\scriptsize,gray!50!black,anchor=north west] at (axis cs:0.3,0.338) {$\rho^\star$};
\end{axis}
\begin{axis}[at={(102.0mm,9mm)},anchor=south west,scale only axis,width=28mm,height=24.0mm,xmin=0,xmax=1,xtick={0,0.5,1},label style={font=\footnotesize},tick label style={font=\footnotesize},x label style={yshift=1.5mm},y label style={at={(axis description cs:-0.3,0.5)}},axis background/.style={fill=white},title style={font=\footnotesize\bfseries,yshift=-1mm},ymin=0.15,ymax=0.34,ytick={0.2,0.25,0.3},yticklabel style={/pgf/number format/fixed},xlabel={dose $\rho=\DA$},title={(c) $\varepsilon=0.5$},legend style={font=\scriptsize,at={(0.97,0.97)},anchor=north east,draw=none,fill=white,fill opacity=0.85,text opacity=1,row sep=-2pt}]
\addplot[black,thick] coordinates {(0.0,0.2214) (0.1,0.2177) (0.2,0.2139) (0.3,0.2102) (0.4,0.2064) (0.5,0.2027) (0.6,0.1989) (0.7,0.1952) (0.8,0.1914) (0.9,0.1877) (1.0,0.1839)};
\addplot[red!80!black,thick,densely dashed,mark=o,mark size=0.9pt] coordinates {(0.0,0.2472) (0.1,0.2405) (0.2,0.2315) (0.3,0.2236) (0.4,0.2154) (0.5,0.2075) (0.6,0.2012) (0.7,0.1920) (0.8,0.1841) (0.9,0.1757) (1.0,0.1676)};
\addplot[blue!80!black,thick,dashdotted,mark=square,mark size=0.8pt] coordinates {(0.0,0.2472) (0.1,0.2442) (0.2,0.2390) (0.3,0.2348) (0.4,0.2304) (0.5,0.2263) (0.6,0.2237) (0.7,0.2183) (0.8,0.2141) (0.9,0.2095) (1.0,0.2051)};
\legend{true rate $p$,plain bound,oracle correction}
\draw[gray,densely dotted,thick] (axis cs:0.3,0.15) -- (axis cs:0.3,0.34);
\node[font=\scriptsize,gray!50!black,anchor=north west] at (axis cs:0.3,0.338) {$\rho^\star$};
\end{axis}
\begin{axis}[at={(147.0mm,9mm)},anchor=south west,scale only axis,width=28mm,height=24.0mm,xmin=0,xmax=1,xtick={0,0.5,1},label style={font=\footnotesize},tick label style={font=\footnotesize},x label style={yshift=1.5mm},y label style={at={(axis description cs:-0.3,0.5)}},axis background/.style={fill=white},title style={font=\footnotesize\bfseries,yshift=-1mm},ymin=0,ymax=1.05,ytick={0,0.5,0.95},yticklabel style={/pgf/number format/fixed},xlabel={dose $\rho=\DA$},title={(d) plain coverage},legend style={font=\scriptsize,at={(0.03,0.03)},anchor=south west,draw=none,fill=white,fill opacity=0.85,text opacity=1,row sep=-2pt}]
\addplot[black,densely dashed,domain=0:1,samples=2,forget plot]{0.95};
\addplot[orange!80!black,thick,mark=*,mark size=0.8pt] coordinates {(0.0,1.0000) (0.1,0.9970) (0.2,0.9770) (0.3,0.9730) (0.4,0.9500) (0.5,0.9530) (0.6,0.9130) (0.7,0.8930) (0.8,0.8200) (0.9,0.7900) (1.0,0.6430)};
\addlegendentry{$\varepsilon=0.05$}
\addplot[blue!80!black,thick,mark=*,mark size=0.8pt] coordinates {(0.0,0.9970) (0.1,0.9970) (0.2,0.9730) (0.3,0.9100) (0.4,0.8500) (0.5,0.7670) (0.6,0.6200) (0.7,0.3870) (0.8,0.2600) (0.9,0.1370) (1.0,0.0370)};
\addlegendentry{$\varepsilon=0.2$}
\addplot[green!50!black,thick,mark=*,mark size=0.8pt] coordinates {(0.0,0.9900) (0.1,0.9800) (0.2,0.9670) (0.3,0.9130) (0.4,0.8130) (0.5,0.7000) (0.6,0.5670) (0.7,0.3630) (0.8,0.2130) (0.9,0.0830) (1.0,0.0470)};
\addlegendentry{$\varepsilon=0.5$}
\end{axis}
\end{tikzpicture}}
\caption{E6 dose--response on the fully measured gpt-4.1-mini population (erasure channel). (a)--(c)~Mean certified bounds against the true rate; upper curve: oracle correction $(1-\varepsilon)\rho\,\kappa_{\mathcal A}$, not Theorem~\ref{thm:inflated}; dotted: $\rho^\star$, where the plain bound's coverage drops below $0.95$. (d)~Coverage of the plain bound; dashed: the $0.95$ target.}
\label{fig:e5}
\end{figure*}

\paragraph*{Dose--response and the audit rate (E6)} Opening the channel at a graded rate $\rho$ sweeps $\DA$ over $[0,1]$ and the exploited gap over $[0,0.075]$ (Fig.~\ref{fig:e5}). Coverage of $p_i$ crosses $0.95$ between $\rho{=}0.2$ and $0.3$ for $\varepsilon{=}0.2$. An oracle correction by the exploited gap, known only from the full labels, stays above the truth; Theorem~\ref{thm:inflated} with $\delta{=}\rho$ is vacuous here. Raising the audit rate from $0.05$ to $0.2$ shrinks the bias, but it shrinks the width faster, so $n^\star$ falls from $4.9{\times}10^3$ to $1.6{\times}10^3$, as Corollary~\ref{cor:eps} predicts below $\varepsilon^\star\approx0.35$; at $\varepsilon{=}0.5$ it stays near $1.5{\times}10^3$. A Hoeffding sequence hides the failure by being loose; the variance-adaptive certifier the theory specifies is the one that breaks.

\subsection{Does it happen in a live run?}

\begin{findbox}{Finding C}
On a pre-registered schedule of $4000$ live episodes the frozen test gave only its second-tier verdict (App.~\ref{app:expdetails}). A sequence valid for the realized rate under PNI, with $\lambda$ fixed at the pre-registered horizon (a post-hoc analysis), drops below it at \bad{$n{=}2687$}, inside the pre-registered $n^\star$ interval $[0.9,3.5]{\times}10^3$, and stays below under the $200$-step rule; the silent arm on the same schedule \good{covers at every $n$}.
\end{findbox}

\paragraph*{A pre-registered failure run (E8)} A first closed loop on the real timing channel (E7, App.~\ref{app:expdetails}; $N{=}500$ per arm) produced a gap that puts $n^\star$ near $3{\times}10^4$, and the certificate covered, as Theorem~\ref{thm:nands}(ii) predicts below $n^\star$. The failure regime needs the widest channel, so E8 uses E3's notice channel with the named clause on the full $560$-pair population. The design was pre-registered (App.~\ref{app:expdetails}). A fixed-seed schedule of pairs and coins ($\varepsilon{=}0.2$) was hashed before the first episode, together with the pair-bootstrap prediction $n^\star\approx1.6{\times}10^3$ $[0.9,3.5]{\times}10^3$ and a persistence rule for the crossing. Both arms share the schedule. The \emph{plain} arm appends the notice to the first tool output on audited episodes; the \emph{silent} arm appends nothing and takes its label on the same episodes. The benchmark judges every episode, so the realized rate $\bar V_n$ is known while the certificate sees only audited labels.

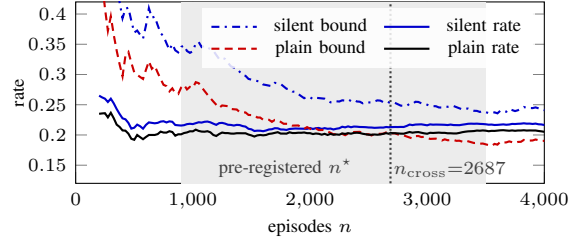
\begin{figure}[t]\centering
% generated by experiments/m13_fig_dose.py -- do not edit by hand
\begin{tikzpicture}[font=\footnotesize]
\begin{axis}[scale only axis,width=62mm,height=24mm,xmin=0,xmax=4000,xtick={0,1000,2000,3000,4000},label style={font=\scriptsize},tick label style={font=\scriptsize},x label style={yshift=1mm},y label style={at={(axis description cs:-0.09,0.5)}},ymin=0.12,ymax=0.42,ytick={0.15,0.2,0.25,0.3,0.35,0.4},ylabel={rate},yticklabel style={/pgf/number format/fixed},xlabel={episodes $n$},legend style={font=\scriptsize,at={(0.98,0.97)},anchor=north east,draw=none,fill=white,fill opacity=0.85,text opacity=1,row sep=-2pt,column sep=4pt},legend columns=2,clip=true]
\fill[gray!15] (axis cs:900,0.12) rectangle (axis cs:3500,0.42);
\addplot[blue!80!black,thick,dashdotted] coordinates {(200,0.6000) (225,0.6000) (250,0.5412) (275,0.4920) (300,0.5242) (325,0.4839) (350,0.4493) (375,0.4194) (400,0.3931) (425,0.4088) (450,0.4104) (475,0.3888) (500,0.3694) (525,0.3727) (550,0.3658) (575,0.3594) (600,0.3810) (625,0.4097) (650,0.3939) (675,0.3956) (700,0.3815) (725,0.3683) (750,0.3634) (775,0.3517) (800,0.3407) (825,0.3503) (850,0.3400) (875,0.3366) (900,0.3455) (925,0.3480) (950,0.3389) (975,0.3358) (1000,0.3384) (1025,0.3516) (1050,0.3537) (1075,0.3454) (1100,0.3476) (1125,0.3447) (1150,0.3420) (1175,0.3394) (1200,0.3323) (1225,0.3255) (1250,0.3278) (1275,0.3257) (1300,0.3194) (1325,0.3134) (1350,0.3116) (1375,0.3100) (1400,0.3044) (1425,0.2991) (1450,0.2977) (1475,0.2927) (1500,0.2915) (1525,0.2903) (1550,0.2892) (1575,0.2880) (1600,0.2835) (1625,0.2826) (1650,0.2816) (1675,0.2807) (1700,0.2862) (1725,0.2853) (1750,0.2843) (1775,0.2803) (1800,0.2764) (1825,0.2757) (1850,0.2719) (1875,0.2712) (1900,0.2677) (1925,0.2642) (1950,0.2608) (1975,0.2631) (2000,0.2625) (2025,0.2593) (2050,0.2615) (2075,0.2610) (2100,0.2605) (2125,0.2600) (2150,0.2570) (2175,0.2540) (2200,0.2536) (2225,0.2557) (2250,0.2553) (2275,0.2573) (2300,0.2569) (2325,0.2589) (2350,0.2561) (2375,0.2580) (2400,0.2554) (2425,0.2550) (2450,0.2524) (2475,0.2543) (2500,0.2539) (2525,0.2536) (2550,0.2554) (2575,0.2572) (2600,0.2568) (2625,0.2544) (2650,0.2561) (2675,0.2537) (2687,0.2546) (2700,0.2534) (2725,0.2531) (2750,0.2508) (2775,0.2485) (2800,0.2483) (2825,0.2480) (2850,0.2516) (2875,0.2514) (2900,0.2549) (2925,0.2546) (2950,0.2524) (2975,0.2521) (3000,0.2500) (3025,0.2480) (3050,0.2477) (3075,0.2475) (3100,0.2455) (3125,0.2453) (3150,0.2468) (3175,0.2466) (3200,0.2481) (3225,0.2479) (3250,0.2460) (3275,0.2441) (3300,0.2423) (3325,0.2437) (3350,0.2436) (3375,0.2418) (3400,0.2400) (3425,0.2398) (3450,0.2381) (3475,0.2395) (3500,0.2378) (3525,0.2361) (3550,0.2360) (3575,0.2374) (3600,0.2358) (3625,0.2402) (3650,0.2386) (3675,0.2384) (3700,0.2428) (3725,0.2412) (3750,0.2395) (3775,0.2409) (3800,0.2436) (3825,0.2463) (3850,0.2447) (3875,0.2446) (3900,0.2444) (3925,0.2456) (3950,0.2441) (3975,0.2425) (4000,0.2410)};
\addplot[blue!80!black,thick] coordinates {(200,0.2650) (225,0.2622) (250,0.2600) (275,0.2545) (300,0.2600) (325,0.2523) (350,0.2429) (375,0.2347) (400,0.2275) (425,0.2259) (450,0.2200) (475,0.2105) (500,0.2120) (525,0.2171) (550,0.2145) (575,0.2104) (600,0.2167) (625,0.2208) (650,0.2200) (675,0.2193) (700,0.2200) (725,0.2221) (750,0.2227) (775,0.2207) (800,0.2225) (825,0.2218) (850,0.2165) (875,0.2160) (900,0.2156) (925,0.2162) (950,0.2179) (975,0.2185) (1000,0.2180) (1025,0.2195) (1050,0.2210) (1075,0.2205) (1100,0.2218) (1125,0.2222) (1150,0.2209) (1175,0.2187) (1200,0.2167) (1225,0.2196) (1250,0.2200) (1275,0.2188) (1300,0.2169) (1325,0.2166) (1350,0.2156) (1375,0.2167) (1400,0.2164) (1425,0.2147) (1450,0.2138) (1475,0.2102) (1500,0.2093) (1525,0.2066) (1550,0.2071) (1575,0.2070) (1600,0.2062) (1625,0.2074) (1650,0.2091) (1675,0.2095) (1700,0.2112) (1725,0.2110) (1750,0.2091) (1775,0.2090) (1800,0.2083) (1825,0.2099) (1850,0.2103) (1875,0.2107) (1900,0.2105) (1925,0.2104) (1950,0.2103) (1975,0.2101) (2000,0.2115) (2025,0.2099) (2050,0.2112) (2075,0.2116) (2100,0.2100) (2125,0.2113) (2150,0.2102) (2175,0.2106) (2200,0.2091) (2225,0.2103) (2250,0.2111) (2275,0.2136) (2300,0.2126) (2325,0.2142) (2350,0.2128) (2375,0.2139) (2400,0.2137) (2425,0.2132) (2450,0.2131) (2475,0.2129) (2500,0.2124) (2525,0.2123) (2550,0.2118) (2575,0.2120) (2600,0.2123) (2625,0.2122) (2650,0.2132) (2675,0.2127) (2687,0.2136) (2700,0.2130) (2725,0.2132) (2750,0.2131) (2775,0.2133) (2800,0.2132) (2825,0.2149) (2850,0.2172) (2875,0.2163) (2900,0.2172) (2925,0.2174) (2950,0.2176) (2975,0.2168) (3000,0.2160) (3025,0.2175) (3050,0.2170) (3075,0.2172) (3100,0.2171) (3125,0.2170) (3150,0.2168) (3175,0.2164) (3200,0.2172) (3225,0.2164) (3250,0.2163) (3275,0.2162) (3300,0.2170) (3325,0.2177) (3350,0.2185) (3375,0.2187) (3400,0.2185) (3425,0.2187) (3450,0.2183) (3475,0.2181) (3500,0.2177) (3525,0.2170) (3550,0.2172) (3575,0.2165) (3600,0.2167) (3625,0.2171) (3650,0.2175) (3675,0.2169) (3700,0.2173) (3725,0.2183) (3750,0.2181) (3775,0.2185) (3800,0.2187) (3825,0.2193) (3850,0.2187) (3875,0.2188) (3900,0.2182) (3925,0.2188) (3950,0.2177) (3975,0.2174) (4000,0.2165)};
\addplot[red!80!black,thick,densely dashed] coordinates {(200,0.5119) (225,0.4550) (250,0.4095) (275,0.3723) (300,0.3961) (325,0.3656) (350,0.3395) (375,0.3169) (400,0.2971) (425,0.3313) (450,0.3373) (475,0.3195) (500,0.3035) (525,0.2995) (550,0.2959) (575,0.2926) (600,0.3078) (625,0.3219) (650,0.3095) (675,0.3143) (700,0.3031) (725,0.2926) (750,0.2829) (775,0.2808) (800,0.2721) (825,0.2771) (850,0.2690) (875,0.2676) (900,0.2784) (925,0.2828) (950,0.2753) (975,0.2795) (1000,0.2780) (1025,0.2873) (1050,0.2857) (1075,0.2791) (1100,0.2727) (1125,0.2666) (1150,0.2609) (1175,0.2553) (1200,0.2500) (1225,0.2449) (1250,0.2488) (1275,0.2482) (1300,0.2476) (1325,0.2430) (1350,0.2425) (1375,0.2421) (1400,0.2378) (1425,0.2375) (1450,0.2372) (1475,0.2331) (1500,0.2329) (1525,0.2327) (1550,0.2289) (1575,0.2288) (1600,0.2252) (1625,0.2251) (1650,0.2251) (1675,0.2217) (1700,0.2249) (1725,0.2248) (1750,0.2216) (1775,0.2185) (1800,0.2185) (1825,0.2185) (1850,0.2155) (1875,0.2156) (1900,0.2157) (1925,0.2129) (1950,0.2101) (1975,0.2130) (2000,0.2104) (2025,0.2078) (2050,0.2106) (2075,0.2081) (2100,0.2082) (2125,0.2083) (2150,0.2059) (2175,0.2035) (2200,0.2012) (2225,0.2014) (2250,0.2016) (2275,0.2018) (2300,0.1996) (2325,0.2022) (2350,0.2024) (2375,0.2049) (2400,0.2050) (2425,0.2052) (2450,0.2031) (2475,0.2010) (2500,0.2012) (2525,0.1992) (2550,0.2016) (2575,0.2018) (2600,0.2019) (2625,0.2021) (2650,0.2043) (2675,0.2024) (2687,0.2036) (2700,0.2026) (2725,0.2027) (2750,0.2009) (2775,0.1991) (2800,0.1993) (2825,0.1975) (2850,0.1958) (2875,0.1941) (2900,0.1981) (2925,0.1983) (2950,0.1966) (2975,0.1968) (3000,0.1951) (3025,0.1935) (3050,0.1937) (3075,0.1939) (3100,0.1924) (3125,0.1908) (3150,0.1928) (3175,0.1913) (3200,0.1915) (3225,0.1917) (3250,0.1903) (3275,0.1905) (3300,0.1890) (3325,0.1909) (3350,0.1895) (3375,0.1881) (3400,0.1867) (3425,0.1870) (3450,0.1856) (3475,0.1858) (3500,0.1845) (3525,0.1832) (3550,0.1819) (3575,0.1852) (3600,0.1840) (3625,0.1888) (3650,0.1875) (3675,0.1877) (3700,0.1894) (3725,0.1881) (3750,0.1868) (3775,0.1885) (3800,0.1887) (3825,0.1918) (3850,0.1905) (3875,0.1907) (3900,0.1909) (3925,0.1925) (3950,0.1913) (3975,0.1901) (4000,0.1903)};
\addplot[black,thick] coordinates {(200,0.2350) (225,0.2356) (250,0.2360) (275,0.2291) (300,0.2367) (325,0.2277) (350,0.2171) (375,0.2133) (400,0.2075) (425,0.2071) (450,0.2022) (475,0.1937) (500,0.1920) (525,0.2000) (550,0.1964) (575,0.1930) (600,0.1967) (625,0.2000) (650,0.2000) (675,0.2000) (700,0.2014) (725,0.2055) (750,0.2040) (775,0.2039) (800,0.2062) (825,0.2049) (850,0.1988) (875,0.2011) (900,0.2011) (925,0.2032) (950,0.2042) (975,0.2061) (1000,0.2060) (1025,0.2088) (1050,0.2067) (1075,0.2056) (1100,0.2054) (1125,0.2062) (1150,0.2044) (1175,0.2000) (1200,0.1983) (1225,0.2016) (1250,0.2040) (1275,0.2031) (1300,0.2031) (1325,0.2023) (1350,0.2022) (1375,0.2029) (1400,0.2029) (1425,0.2028) (1450,0.2028) (1475,0.2007) (1500,0.1993) (1525,0.1974) (1550,0.1993) (1575,0.2000) (1600,0.1994) (1625,0.2000) (1650,0.2018) (1675,0.2024) (1700,0.2035) (1725,0.2046) (1750,0.2023) (1775,0.2023) (1800,0.2022) (1825,0.2044) (1850,0.2032) (1875,0.2032) (1900,0.2037) (1925,0.2036) (1950,0.2041) (1975,0.2030) (2000,0.2030) (2025,0.2015) (2050,0.2024) (2075,0.2019) (2100,0.2005) (2125,0.2019) (2150,0.2000) (2175,0.2018) (2200,0.2004) (2225,0.2009) (2250,0.2004) (2275,0.2026) (2300,0.2017) (2325,0.2030) (2350,0.2021) (2375,0.2034) (2400,0.2037) (2425,0.2033) (2450,0.2020) (2475,0.2012) (2500,0.2004) (2525,0.2004) (2550,0.2008) (2575,0.2004) (2600,0.2008) (2625,0.2019) (2650,0.2030) (2675,0.2019) (2687,0.2036) (2700,0.2030) (2725,0.2029) (2750,0.2026) (2775,0.2022) (2800,0.2021) (2825,0.2035) (2850,0.2049) (2875,0.2038) (2900,0.2045) (2925,0.2048) (2950,0.2037) (2975,0.2027) (3000,0.2020) (3025,0.2036) (3050,0.2033) (3075,0.2036) (3100,0.2039) (3125,0.2035) (3150,0.2035) (3175,0.2028) (3200,0.2028) (3225,0.2025) (3250,0.2025) (3275,0.2034) (3300,0.2046) (3325,0.2054) (3350,0.2060) (3375,0.2062) (3400,0.2065) (3425,0.2073) (3450,0.2070) (3475,0.2075) (3500,0.2071) (3525,0.2062) (3550,0.2068) (3575,0.2067) (3600,0.2072) (3625,0.2077) (3650,0.2079) (3675,0.2074) (3700,0.2070) (3725,0.2075) (3750,0.2069) (3775,0.2074) (3800,0.2074) (3825,0.2078) (3850,0.2073) (3875,0.2075) (3900,0.2074) (3925,0.2074) (3950,0.2066) (3975,0.2060) (4000,0.2052)};
\legend{silent bound,silent rate,plain bound,plain rate}
\draw[gray!50!black,densely dotted,thick] (axis cs:2687,0.12) -- (axis cs:2687,0.42);
\node[font=\scriptsize,gray!50!black,anchor=south west,inner sep=1pt] at (axis cs:2687,0.125) {$n_{\mathrm{cross}}{=}2687$};
\node[font=\scriptsize,gray!50!black,anchor=south,inner sep=1pt] at (axis cs:1793.5,0.125) {pre-registered $n^\star$};
\end{axis}
\end{tikzpicture}
\caption{E8 live run (gpt-4.1-mini, notice channel, $\varepsilon{=}0.2$): realized rate $\bar V_n$ and realized-rate bound (Prop.~\ref{prop:realized}, $\lambda$ fixed at the pre-registered horizon; a post-hoc analysis; the pre-registered comparison crosses at $2818$) per arm; shaded: pre-registered $n^\star$ interval.}
\label{fig:e8}
\end{figure}

\paragraph*{Result} The deployed sequence certifies $\frac1n\sum_ip_i$, which no run reveals. Fig.~\ref{fig:e8} therefore compares $\bar V_n$ with the realized-rate sequence of Proposition~\ref{prop:realized}, which is valid for $\bar V_n$ under PNI; its $\lambda$ was fixed from the pre-registered planning rate, using no outcomes. Over $4000$ plain episodes and $776$ audits the plain bound, with $\lambda$ fixed at the pre-registered horizon $3000$, first dips below $\bar V_n$ at $n{=}2206$, stays below from $n{=}2687$, inside the $n^\star$ interval, and at $N$ certifies $0.190$ against a realized $0.205$; the silent arm covers at every $n$ (re-tuning $\lambda$ to $4000$ gives $2746$ and $0.191$). The two arms share the schedule, so the $776$ audited episodes pair across arms with the notice as the only difference: the silent arm attacks on $54$ of them where the plain arm does not, against $17$ the other way. The $776$ episodes come from $423$ distinct pairs, so the test is over pairs ($43$ vs.\ $13$, sign test $p{=}7{\times}10^{-5}$), a within-run estimate $\kappa_{\mathcal A}{=}0.048$ $[0.025,0.071]$ (pair bootstrap) independent of the population bootstrap. A certifier valid by construction under PNI and (ER) fails on real labels once the coin reaches the view, which is what \S\ref{sec:theory} predicted and \S\ref{sec:mechanism} prevents.

\paragraph*{Deviations and scope} Two deviations are recorded in App.~\ref{app:expdetails}. The pre-registered rule compared the deployed sequence with $\bar V_n$, a target mismatch; under it the crossing is $2818$, and at the frozen primary horizon $N{=}3000$ that rule returned its second-tier verdict, after which the run was extended to the design's listed alternative of $4000$. Episodes are stateless draws from the $560$-pair population, so cross-episode adaptation is not tested.

%====================================================================
\section{Discussion and Limitations}
\label{sec:disc}

\paragraph*{What the guarantee covers} We analyze the dangerous case $\mu_1<\mu_0$, in which the agent attacks less when it infers that it is audited. In the opposite case the certificate merely over-covers, and the bound $(1-\varepsilon)\DA$ holds regardless of sign. A violation concealed even from the auditor has no label and needs a prevention argument.

\paragraph*{Adversary-controlled quantities} Any quantity the adversary controls must enter the certificate in one direction only. An agent could inflate its ingest on the episodes it means to violate, so that they are excluded as exceedances; this is the selective-labels failure. Fail-closed counting defuses it, and the measured-$\bar G$ readout or the cap defuses call-count inflation. The residual exposure is availability, detectable and never anti-conservative.

\paragraph*{Shared infrastructure} A neighbour that learns another episode's coin is harmless, since the certificate needs only $V_j\perp B_j$. The dangerous path is an \emph{aggregate}. Under concurrent batch processing, audited episodes raise a shared, agent-readable load statistic to which an episode's own coin contributes. Proposition~\ref{prop:parallel} computes this leakage exactly for a load statistic that counts audited episodes: the mode probability of $\mathrm{Bin}(M{-}1,\varepsilon)$, nonzero for every $M$ and decaying only as $M^{-1/2}$. On the real gate with $M$ from $2$ to $20$ batch-synchronous workers, the measured lower bounds sit below the noiseless values and decay the same way (App.~\ref{app:assess}). Per-call coins do not help; core pinning only partly does. Closing it needs isolation, $B$-independent cover load, or post-commit placement.

\paragraph*{What a practitioner gets} Each step of the recipe has a price in width. With a post-hoc audit or a silent branch the certificate is the plain sequence; on E5's population it certifies $0.265$ against $\mu_0{=}0.221$ (deployed rate $0.206$). A per-call coin multiplies the certified value by $\E[v\mid\text{violating}]$, at least $1.7$ on the same population by the goal-count proxy. A residual $\delta$ adds $(1-\varepsilon)\delta$, about $0.05$ for the coarsened stamps (a diagnostic figure, App.~\ref{app:simul}). Below a base rate of about $0.05$ only the multiplicative form serves. Hence the order of the recipe: prefer the constructions with $\DA{=}0$, pay the readout factor only when a pre-commit branch must run, and pay $\delta$ last.

\paragraph*{Limitations} \emph{Stacks.} The main measured stack is a monitor we built, and its timing channel follows from its design choice to redo work in the audit branch before commit; the second stack is thin. A complete run on a public monitor we did not build is needed before the channel can be called typical. Our stamps are in-process (a remote service adds jitter), and the user-facing cost of decoys is not measured. \emph{Agents.} Real-LLM exploitation is established for an explicit audit-conditioned injection on gpt-4.1-mini; gpt-4o-mini moves the same way, and the spontaneous, generic, and real-agent timing effects are only suggestive. \emph{Scale.} Large-$n$ coverage resamples $560$ real pairs, and E8's live episodes are stateless draws from the same pairs, so cross-episode adaptation, which the threat model admits, is untested and all measured exploitation is within one episode. E8 is one run (one schedule) on one model over the content channel only; a live timing-channel failure remains scripted (E2), since E7's gap puts $n^\star$ near $3{\times}10^4$. \emph{Bounds.} A usable $\delta_\eta$ needs more than $1000$ pairs. We price coin leakage only; missed labels would enter as a second additive term.

\paragraph*{Ethics} The exploitation experiments evaluate a certifier in a local benchmark, with no deployed system and no new model capability. We deliberately do \emph{not} build an instruction that probes timestamps across calls and conceals the attempt, a reusable covert-evasion capability. The generic-clause condition and the pilot end their clause with ``do not mention it''; that text ships in the artifact for reproducibility only.

%====================================================================
\section{Related Work}
\label{sec:related}

\paragraph*{Random auditing of AI systems} \emph{AI control}~\cite{greenblatt2024control} keeps an untrusted model from knowing which outputs are audited. Gans and Holden~\cite{gans2026oversight} show that randomized oversight must be unlearnable before action for \emph{deterrence} to work against a utility-maximizing agent, the setting of inspection games~\cite{avenhaus2002inspection}. Evaluation awareness and sandbagging~\cite{needham2025evalaware} make the point qualitatively, and Santos-Grueiro~\cite{santos2026regime} bounds the evaluation--deployment behaviour gap by mutual information. That observation changes behaviour, and that evaluation context leaks as an information flow, are thus known. What is new is their combination with a sampled-audit certificate: its exact bias price, the identified set, and the commit-time release point (App.~\ref{app:position} compares and marks the classical tools).

\paragraph*{Certification for agents} Conformal methods~\cite{angelopoulos2024crc} freeze a threshold offline; the audit-based line reweights sampled labels by inverse propensity~\cite{xu2024arcps,waudbysmith2024betting} and alone has a deployment-time coin that can leak. Active anytime-valid risk control~\cite{xu2024arcps} queries a label after each sample arrives, from behaviour the query cannot change, and its theorems stand; the new condition arises when the agent can still change it after the draw.

\paragraph*{Information flow and quantitative information flow} Noninterference~\cite{goguen1982} and its declassification refinements~\cite{askarov2007gradual,sabelfeld2009declass} protect \emph{data}; nondeducibility on strategies and interactive noninterference~\cite{wittbold1990,oneill2006interactive} quantify, as PNI does, over a low observer who chooses its inputs adaptively; agent IFC systems~\cite{debenedetti2025camel,costa2025fides} protect the user's data or the integrity of decisions. Quantitative information flow~\cite{smith2009qif,alvim2012gleakage,alvim2020science} measures leakage: its additive capacities~\cite{alvim2014additive} are the form of Theorem~\ref{prop:gleak}, adaptive multi-step leakage~\cite{kopf2007adaptive,boreale2015adaptive} precedes Theorem~\ref{thm:decomp}, and channel compositionality~\cite{kawamoto2017compositionality} frames Propositions~\ref{prop:sem}--\ref{prop:parallel}. None of these treats the \emph{monitor's own state} as the protected secret (\S\ref{sec:model}). Our mechanisms have known counterparts. The silent branch is a two-run secure multi-execution~\cite{devriese2010sme}: the audited run computes the label, and the unaudited run produces every output the agent sees. Padding to a ceiling is Agat's transformation~\cite{agat2000timing} applied to one branch, and its goal, that timing not depend on the secret, is the constant-time discipline verified at system level by Barthe et al.~\cite{barthe2014ct}. Coarse release is the single-step case of predictive mitigation~\cite{askarov2010predictive}, whose leakage budget is our $\delta$.

\paragraph*{Secret monitor state in security} Monitors that can be sensed are evaded: intrusion detection through ambiguities~\cite{ptacek1998evasion}, sandboxes and honeypots by malware that detects them~\cite{holz2005honeypot}, the threat transparent analysis~\cite{dinaburg2008ether} answers. These designs catch cheating rather than estimate a rate; risk-limiting audits~\cite{lindeman2012rla} estimate, but over ballots fixed before sampling. Parallels outside security are in App.~\ref{app:position}.

%====================================================================
\section{Conclusion}
\label{sec:concl}

Audit-based certification needs the audit coin kept out of the agent's view before commit. For the episode coin PNI characterizes the unbiased certificate exactly, and its violation costs $(1-\varepsilon)\DA$. On the stacks we measured the leakage is free and agent-controlled, and an agent told what to look for makes the plain certificate fail live; a silent kernel or a per-call coin restores validity.

\vspace{3pt}\noindent\emph{Use of generative AI.} Generative AI tools assisted with copy-editing, literature screening, and code review; the authors verified every claim, number, and reference.

\clearpage
\bibliographystyle{IEEEtran}
\bibliography{refs}

%====================================================================
\clearpage
\appendices

\begin{table*}[tp]\centering\small
\caption{Experiment index (API: needs model calls; script names follow an earlier internal numbering).}
\label{tab:index}
\setlength{\tabcolsep}{3pt}\footnotesize
\begin{tabular}{@{}l>{\raggedright\arraybackslash}p{0.30\textwidth}c>{\raggedright\arraybackslash}p{0.15\textwidth}>{\raggedright\arraybackslash}p{0.33\textwidth}@{}}
\toprule
ID & what it establishes & API & scale & scripts\\
\midrule
E1 & content/timing channels; per-path bounds; padded timestamp observer & no & $2{\times}1000$ pairs; $3000$ calib.\ runs & \texttt{e2\_paired\_v2}, \texttt{m7\_padded} (e2), \texttt{r8\_perpath}\\
E1-ts & timestamp channel, $\DA\ge0.99$ (two runs); coarsened stamps; $E_\eta$ & no & $2{\times}1000$ pairs & \texttt{a6\_timestamp\_clock}, \texttt{m7\_padded} (ts), \texttt{m11\_coarsen}, \texttt{m12\_hockey}\\
E1-amp & $\DA$ agent-inflatable ($\le0.044{\to}{\ge}0.91$) & no & $5{\times}80$ pairs ($5{\times}60$ replay not used) & \texttt{inputsize\_padding} (fixed ceiling), \texttt{a3\_closure\_amplification}\\
E1-isp & size-aware padding; bytes; simultaneous $\delta$ & no & $5{\times}100$ calib.; $5{\times}80$; $2{\times}5{\times}500$; bytes $10{\times}60{+}10{\times}100$ & \texttt{inputsize\_padding}, \texttt{isp\_rerun}, \texttt{m7\_padded} (grid), \texttt{m10\_bytesize}\\
E1-load & padding under load; exceedance counting & no & $3000{+}3{\times}500$ & \texttt{a2\_exceedance\_under\_load}, \texttt{a4\_stress}, \texttt{isp\_load}\\
E1-agg & aggregate-load channel & no & $8{\times}300$ rounds & \texttt{m10\_aggload}\\
CaMeL & timing channel on a second stack & yes & $128$ checks & \texttt{camel\_leak}, \texttt{camel\_c2st}\\
E2 & scripted break; clock and per-call coin & no & $4000{\times}200$ seeds, $k{\le}10^5$ & \texttt{a1\_e6\_redo}\\
E3 & real LLM follows a notice-naming injection & yes & $560{+}816$ ep. & \texttt{e3\_screen}, \texttt{e3\_pilot}, \texttt{e3\_analyze}\\
E3g & generic conditioning, two models & yes & $75{+}394$ ep. & \texttt{e3\_generic}\\
E4 & timing feasibility probe; capability probe & yes & $272$ ep. & \texttt{e3\_timing}, \texttt{timestamp\_capability}\\
nonvuln & full population, measured & yes & $424$ pairs & \texttt{e3\_nonvuln}, \texttt{population}\\
retest & screening noise & yes & $150$ ep. & \texttt{m10\_retest}\\
E5 & plain vs.\ silent, $2{\times}2$; widths; decoys; resampled; pair bootstrap & yes & $400$ seeds; $4000$ boot. & \texttt{e4\_endtoend}, \texttt{e4\_widths}, \texttt{decoy\_sim} (\texttt{POP=measured}), \texttt{pair\_bootstrap}\\
E6 & dose--response; certifier choice; analytic check; resampled & part & $300$ seeds & \texttt{e5\_dose\_response}, \texttt{e5\_reanalyze}, \texttt{r5\_analytic}, \texttt{m13\_fig\_dose}\\
Adapt. & deterministic/plaintext/PRF rules & no & simulation & \texttt{adaptive\_rule\_sim}\\
E7 & live closed loop & yes & $2{\times}500$ ep. & \texttt{e2e\_live}\\
E8 & live failure run, pre-registered; realized-rate re-analysis & yes & $2{\times}4000$ ep. & \texttt{e8\_live\_failure}\\
Dil. & dilution readouts (App.~\ref{app:dilute}) & no & $4{\times}200$ seeds & \texttt{dilute\_sim}\\
Cost & per-call readout looseness on E8 (App.~\ref{app:dilute}) & no & E8 labels & \texttt{m15\_percall\_cost}\\
Aux. & running example; oracle check; KL table; $c_{\mathrm{rng}}$; E4 MDE & no & simulation & \texttt{m14\_constructed}\\
\bottomrule
\end{tabular}
\end{table*}
\section{Positioning and provenance}
\label{app:position}
\begin{table*}[t]\centering\footnotesize
\caption{Comparison with the closest lines of work.}
\label{tab:compare}
\setlength{\tabcolsep}{4pt}
\begin{tabular}{@{}>{\raggedright\arraybackslash}p{0.2\textwidth}>{\raggedright\arraybackslash}p{0.15\textwidth}>{\raggedright\arraybackslash}p{0.13\textwidth}>{\raggedright\arraybackslash}p{0.16\textwidth}>{\raggedright\arraybackslash}p{0.14\textwidth}>{\raggedright\arraybackslash}p{0.1\textwidth}@{}}
\toprule
\textbf{Work} & \textbf{Secret} & \textbf{Fixed when drawn?} & \textbf{Adversary} & \textbf{Estimand} & \textbf{Sharp price}\\
\midrule
Ringers, PoR~\cite{golle2001uncheatable,juels2007por} & which item is checked & yes (work or data) & cheating worker/server & detection & n/a\\
Risk-limiting audits~\cite{lindeman2012rla} & none & yes (ballots) & none adaptive & outcome & n/a\\
Active risk control~\cite{xu2024arcps} & none needed & yes (query after arrival) & not reactive to the query & risk & n/a\\
AI control~\cite{greenblatt2024control} & which outputs are audited & no & untrusted model & safety of protocol & no\\
Randomized oversight~\cite{gans2026oversight} & audit draw & no & utility maximizer & deterrence & no\\
Regime leakage~\cite{santos2026regime} & evaluation vs.\ deployment & no & evaluated model & behaviour gap & MI upper bound\\
\textbf{This work} & audit coin $B_\tau$ & no, until commit & any adaptive policy, Kerckhoffs & $p_i$, $\mu_0$, realized rate & yes (Thm.~\ref{prop:sharp})\\
\bottomrule
\end{tabular}
\end{table*}

Table~\ref{tab:compare} compares the closest lines of work on five axes: what is secret, whether the sampled behaviour is already fixed when the draw is made, the adversary, the estimand, and whether a sharp price of leakage is given. Table~\ref{tab:provenance} separates the classical tools we apply from the statements we claim as new.

\begin{table}[h]\centering\footnotesize
\caption{Provenance of the formal results.}
\label{tab:provenance}
\setlength{\tabcolsep}{2pt}
\begin{tabular}{@{}>{\raggedright\arraybackslash}p{0.27\columnwidth}>{\raggedright\arraybackslash}p{0.69\columnwidth}@{}}
\toprule
\textbf{Result} & \textbf{Classical tool applied / what is new}\\
\midrule
Lemmas~\ref{lem:identity},~\ref{lem:bridge} & IPW identity; variational form of total variation (Neyman--Pearson). New: the gap as a test of the pre-commit view.\\
Thm.~\ref{prop:gleak} & QIF vulnerability, additive capacity, post-processing~\cite{alvim2020science,alvim2014additive}. Instantiation: the gain function $g_\varepsilon$ for which the bias \emph{is} the leakage.\\
Thm.~\ref{thm:valid}, Prop.~\ref{prop:realized} & Time-uniform sequences~\cite{howard2021timeuniform}, Bennett, Ville. New: the conditions (PNI, ER) under which they apply to audit labels.\\
Thm.~\ref{thm:decomp}, Prop.~\ref{prop:kl} & Coupling inequality~\cite{lindvall1992coupling}; KL chain rule and Pinsker; adaptive leakage after~\cite{kopf2007adaptive}. Applied to coin-to-view channels.\\
Thm.~\ref{thm:nands} & New: PNI is necessary and sufficient; coverage fails under persistent leakage.\\
Prop.~\ref{prop:prf} & Standard PRF hybrid~\cite{goldreich1986prf}. New: the reduction to the audit game, with the per-history conditioning cost.\\
Thm.~\ref{thm:inflated}, Thm.~\ref{prop:sharp} & Partial identification~\cite{manski2003partial}. New: the inflation and its sharpness from audited labels.\\
Prop.~\ref{prop:hockey}, Cor.~\ref{cor:blackwell} & Hockey-stick divergence; Blackwell order~\cite{blackwell1953}. Instantiation for the audit channel.\\
Prop.~\ref{prop:sem} & New: the constructions; noninterference, multi-execution, and QIF compositionality~\cite{goguen1982,devriese2010sme,kawamoto2017compositionality} as background.\\
\bottomrule
\end{tabular}
\end{table}

\section{Experiment index and reproduction}
\label{app:repro}
Table~\ref{tab:index} maps each claim to its scale, whether it uses model APIs, and its scripts in \texttt{experiments/}, the authoritative map from experiment IDs to file names. API-free experiments and all re-analyses are deterministic given the recorded latencies and labels; the paired design records the potential outcomes $(V(Z{=}0),V(Z{=}1))$ per (user task, injection task), so E5/E6 coverage is recomputed offline from real labels. Model experiments use AgentDojo v1.2.2 at temperature $0$ with \texttt{gpt-4o-mini-2024-07-18}, \texttt{gpt-4.1-mini}, \texttt{glm-4.7-flashx}, and \texttt{glm-4.6}; total model-API cost was about US\$91, of which E8 was US\$59. The artifact's README records run configurations, calibration procedures, and the two incidents (one crashed byte-size phase, the glm-4.6 rate-limit errors); \texttt{e2\_tableII.py} prints every row of Table~\ref{tab:e2} from the stored summaries (medians, mismatches, and C2ST also recompute from the raw per-pair file), and the PQ-CIFC stack ships with the artifact. The artifact will be made publicly available upon acceptance.

\section{The measured stack (PQ-CIFC) and the audit branch}
\label{app:stack}
PQ-CIFC is the pre-commit IFC gate of this work, built on AgentDojo~\cite{debenedetti2024agentdojo}. It serves as a concrete monitor whose audit branch we can instrument; none of our results depend on its labeling accuracy. Labels pair a confidentiality level (Public/Internal/Confidential/Secret) with an integrity level (Untrusted/Verified/Trusted); integrity comes from provenance, and the scorer predicts confidentiality only. Each tool output is ingested as a labeled object. Each effect tool call goes through a monitor. The monitor computes the dependency closure of the call and scores it with a trained joint label predictor (TF-IDF features with logistic regression). A broker then commits the effect only if the predicted label flow complies with the policy. We use its calibrated, frozen \emph{quotient} policy (threshold $q{=}0.9525$); under it, the scorer pools each object's class probabilities into ``admissible at this sink'' versus not, takes the minimum over the closure, and forms the conformal set $\{z : 1-\hat p(z)\le q\}$; the effect commits only if the set is the singleton \{admissible\}. Its \emph{full}-label policy calibrates to $q{=}1$ and denies every effect, so it cannot exercise a permitted path. The permitted paths replay the ground-truth calls of AgentDojo workspace user tasks 4, 7, and 8: reads, then \texttt{create\_calendar\_event}, \texttt{reschedule\_calendar\_event}, and \texttt{add\_calendar\_event\_participants}, all allowed and committed. The denied paths ingest a security e-mail and then attempt an external e-mail, an internal e-mail, or a file creation. The effect path is identical for $B{=}0$ and $B{=}1$. Audited episodes additionally run, before the effect, a silent branch that recomputes the closure over all ingested objects, scores it with the same scorer, and writes one row to a SQLite audit log with a synchronous commit. The three parts were not timed separately, so part of the gap may be logging rather than recomputation. The channel comes from this re-done work on audited episodes, not from PQ-CIFC's internals; the CaMeL assessment of \S\ref{sec:sim} shows the same profile.

\emph{Ordering.} The confined gate pads before the effect commits: monitor decision $\to$ (audited: silent branch) $\to$ pad to ceiling $\to$ effect commit (the write that stamps \texttt{last\_modified}) $\to$ response, so the inter-timestamp gap the timestamp observer reads contains the ceiling, not the raw latency; the pad window also absorbs the audit's deferred work, and the exit and commit run with the garbage collector disabled. The legacy harness padded \emph{after} the commit; its exceedance counts on denied paths are unaffected, but its timestamp channel was open (C2ST $0.51$ $[0.47,0.56]$). All padded numbers in \S\ref{sec:sim} come from the confined implementation except the counterfactual count $0.032$ of App.~\ref{app:simul}(iv).

\section{Deferred proofs}
\label{app:proofs}
\emph{Proof of Lemma~\ref{lem:bridge}.} Given $h$, the attempt is a test $\phi(W)\in[0,1]$ of the view, so $\mu_0-\mu_1=\E_{P_0}\phi-\E_{P_1}\phi\le\sup_{0\le\phi\le1}(\E_{P_0}\phi-\E_{P_1}\phi)=\TV(P_0,P_1)\le\DA$ by the variational characterization of total variation (Neyman--Pearson); the supremum is attained by the stated rule. Changing only the decision at $\tau$, which is taken after $W$ is formed, changes neither the law of $W$ nor the law of $H_{i-1}$. The other two terms follow from $\mu_0,\mu_1\in[0,1]$. \hfill$\square$

\emph{Proof that $\Lambda=(1-\varepsilon)\DA$ for the episode coin.} Let $\bar P:=\varepsilon P_1+(1-\varepsilon)P_0$ be the law of $W$. Then $r(W)=\Pr(B{=}1\mid W)/\varepsilon=\mathrm dP_1/\mathrm d\bar P\,(W)$, so $\Lambda=\E_{\bar P}[(1-\mathrm dP_1/\mathrm d\bar P)_+]=\TV(\bar P,P_1)=(1-\varepsilon)\TV(P_0,P_1)$, because $\bar P-P_1=(1-\varepsilon)(P_0-P_1)$. \hfill$\square$

\emph{Two sources of leakage.} For an adaptive rule write $\pi^{\mathrm{true}}:=\Pr_{\mathcal A}(B_\tau{=}1\mid H_{i-1}{=}h)$ and $\rho(w):=\Pr_{\mathcal A}(B_\tau{=}1\mid W{=}w)/\pi^{\mathrm{true}}$, so that $r=\rho\,\pi^{\mathrm{true}}/\pi_\tau$ and $\E[\rho]=1$. Then $\E|1-r|\le\E|1-\rho|+|1-\pi^{\mathrm{true}}/\pi_\tau|\le2(1-\pi^{\mathrm{true}})\DA+|\pi^{\mathrm{true}}-\pi_\tau|/\pi_\tau$, where the first term is the view leakage of Definition~\ref{def:da} and the second a marginal miscalibration of the coin that $\DA$ alone does not see; $\E|1-r|=0$ is unbiasedness, $\Lambda=0$ only rules out under-certification. A keyed rule (Proposition~\ref{prop:prf}) removes the second term; the constructions of Proposition~\ref{prop:sem} remove the first.

\emph{Proof of Theorem~\ref{prop:gleak}.} (i) The expected gain of $\phi$ is $(1-\varepsilon)\E_{P_0}\phi+(1-\varepsilon)(1-\E_{P_1}\phi)=(1-\varepsilon)+(1-\varepsilon)(\mu_0-\mu_1)$; a constant $\phi$ gives the prior vulnerability and the supremum over $\phi$ the posterior one, $(1-\varepsilon)(1+\DA)$ by Lemma~\ref{lem:bridge}. (ii) $\sum_w\max(P_0,P_1)=1+\TV(P_0,P_1)$. (iii) Post-processing increases neither $g$-vulnerability~\cite{alvim2020science} nor total variation. \hfill$\square$

\emph{Proof of Theorem~\ref{thm:valid}.} Under (ER), $V_i$ is a function of the view $W$ and $H_{i-1}$. PNI gives $B_i\perp W\mid H_{i-1}$ jointly over $(O,\xi)$, so $V_i\perp B_i\mid H_{i-1}$. (Independence from $O$ and from $\xi$ separately would not suffice, since $O$ could reveal $B\oplus\xi$.) Hence $\mu_0=\mu_1=p_i$, and Lemma~\ref{lem:identity} gives $\E[\widehat g_i\mid H_{i-1}]=p_i$. Cross-episode learning enters only through $H_{i-1}$, because $B_i$ is a fresh draw. The differences $\widehat g_i-p_i$ are then bounded martingale differences, and the time-uniform bound of~\cite{howard2021timeuniform} controls their running average. For an adaptive rule the same argument applies to $\widehat g=VB_\tau/\pi_\tau(h_\tau)\le1/\pi_{\min}$: the freshness clause of Definition~\ref{def:pni} gives $\E[VB_\tau\mid W,H_{i-1}]=V\pi_\tau(h_\tau)$, hence $\E[\widehat g\mid H_{i-1}]=p_i$. The agent can steer toward low-propensity histories, which widens the sequence through the range $1/\pi_{\min}$, but cannot bias it. \hfill$\square$

\emph{Proof of Corollary~\ref{cor:eps}.} The first claims are Lemma~\ref{lem:identity} and the width formula of App.~\ref{app:width}. For the last, $n^\star\propto f(\varepsilon)=(1/\varepsilon-\bar p)/(1-\varepsilon)^2$, and $f'(\varepsilon)\le0$ iff $2\bar p\varepsilon^2-3\varepsilon+1\ge0$, i.e.\ $\varepsilon\le\varepsilon^\star$. The root $\varepsilon^\star$ increases in $\bar p$ from its limit $\frac13$ at $\bar p\to0$. \hfill$\square$

\emph{Proof of Theorem~\ref{thm:decomp}.} \emph{Construction.} Fix $H_{i-1}$ and the agent's randomness $\xi$, and build the two coin-conditional trajectories on one probability space. Start from the empty common history. At step $t$ with common history $h_t$, the agent's action $a_t$ is the same under both coins, since it is a function of $(h_t,\xi)$. Draw the pair $(S^0_t,S^1_t)$ from the maximal coupling~\cite{lindvall1992coupling} of $P(S_t\mid B{=}0,h_t)$ and $P(S_t\mid B{=}1,h_t)$, under which $S^0_t\ne S^1_t$ with probability exactly $D_t(h_t,a_t)$ and, conditionally on agreement, $S_t$ has the law $\min(P_0,P_1)/(1-D_t)$; the common-history process is therefore the Markov chain with this normalized overlap kernel, stopped at separation, and $\Pr(\sigma\ge\tau)=\E_{\mathcal C}\prod_{t<\tau}(1-D_t)$ is the product of its one-step survival probabilities, by the chain rule along that chain. If they agree, append $(a_t,S_t)$ to the common history; otherwise declare the coupling \emph{broken} at $\sigma:=t$ and continue the two trajectories independently from their own conditionals. Each trajectory has the correct marginal law, so by the coupling inequality $\TV(P^\tau_0,P^\tau_1)\le\Pr(\sigma<\tau)$. Conditional on no break before $t$, the coupling survives step $t$ with probability $1-D_t(h_t,a_t)$, so $\Pr(\sigma\ge\tau)=\E_{\mathcal C}\prod_{t<\tau}(1-D_t(h_t,a_t))$, where $\E_{\mathcal C}$ is over the common history $(h_t)_{t<\sigma\wedge\tau}$; this is the middle term, and Lemma~\ref{lem:bridge} converts it to bias. Replacing $D_t(h,a)$ by $\bar D_t$ and $1-\prod(1-x_t)\le\sum x_t$ give the outer bounds. Under erasure the two laws are mutually singular exactly on the event that some step reveals. \hfill$\square$

\emph{Proof of Proposition~\ref{prop:kl}.} The chain rule for KL divergence over the stopped trajectory gives $\mathrm{KL}(P_0^\tau\|P_1^\tau)\le\sum_{t<T}\bar\kappa_t$, and Pinsker's inequality converts KL to total variation; Lemma~\ref{lem:bridge} converts total variation to bias. \hfill$\square$

\emph{Proof of Theorem~\ref{thm:nands}.} (i) Sufficiency is Theorem~\ref{thm:valid}. For necessity, if $\DA(h;\mathcal A)>0$ for some policy and some $h$ reachable under it, the policy $\mathcal A^{\mathrm{LR}}$ of Lemma~\ref{lem:bridge} reaches the same $h$ and attains $\gapz(h)=\DA(h;\mathcal A)>0$. For an adaptive rule the test $V=\mathbf 1\{r(W)<1\}$ attains $\gapp=\Lambda(h;\mathcal A)$, and $V=\mathbf 1\{r(W)>1\}$ attains $\gapp=-\E[(r-1)_+]$; both vanish for every policy exactly when $r=1$ a.s., and the first vanishes exactly when $\Lambda=0$. A fresh coin with a history-dependent design propensity has $r=1$ yet $\mu_0\ne p$, which is why $\gapz$ is reserved for the episode coin. (ii) The assumption is stated on the histories $\mathcal A^{\mathrm{LR}}$ itself reaches, so Lemma~\ref{lem:bridge} applies in every episode and gives $\E[\widehat g_i\mid H_{i-1}]\le p_i-(1-\varepsilon)D$. By the martingale strong law, the running mean of $\widehat g_i$ then ends up at least $(1-\varepsilon)D$ below that of $p_i$ almost surely, while the certificate's excess tends to $u_\infty<(1-\varepsilon)D$. \hfill$\square$

\emph{Proof of Theorem~\ref{thm:inflated}.} By Lemmas~\ref{lem:identity} and~\ref{lem:bridge}, $\E[\widehat g_i\mid H_{i-1}]\ge p_i-(1-\varepsilon)\delta$ almost surely, so $\frac1n\sum_ip_i\le\frac1n\sum_i\E[\widehat g_i\mid H_{i-1}]+(1-\varepsilon)\delta$. The sequence of~\cite{howard2021timeuniform} upper-bounds the first term for all $n$ at once, and adding a constant keeps it anytime-valid. In the multi-step case take $\delta=1-\prod_t(1-\bar\delta_t)$ by Theorem~\ref{thm:decomp}, where $\bar\delta_t\ge\bar D_t$. \hfill$\square$

\emph{Proof of Theorem~\ref{prop:sharp}.} Lemma~\ref{lem:bridge} gives the upper endpoint as a bound. It is attained: let the view reveal $B$ with probability $t\le\delta$ and otherwise show $\bot$, and let the agent attempt on $0$, skip on $1$, and attempt with probability $c$ on $\bot$. Then $\mu_1=(1-t)c$ and $\mu_0=t+(1-t)c$, so every pair with $0\le\mu_0-\mu_1\le\min(\delta,1-\mu_1)$ is realized by some $(t,c)$, and all these mechanisms induce the same law of $(B,VB)$, so no audited-label procedure separates them. The lower endpoint is symmetric. \hfill$\square$

\emph{Proof of Proposition~\ref{prop:prf}.} A distinguisher with oracle access to $F_K$ or to a random function runs all $N$ episodes of the audit game with $\mathcal A$, simulating the gate, tools, and label oracle, at total cost $N(s+s_M)$. Because every oracle input carries the nonce and the call index, the inputs are distinct, so in the ideal world every coin is a fresh draw. (i) The distinguisher holds every $V_i$ and every coin, so it can output the indicator that the sequence of Proposition~\ref{prop:realized} fails at some $n\le N$. Under $M_{\mathrm{ideal}}$ this event has probability at most $\alpha$, since PNI holds there; its probability under $M_{\mathrm{PRF}}$ is therefore at most $\alpha+\mathrm{Adv}$. (ii) Let the distinguisher output the indicator of $\{H_{i-1}{=}h,V{=}1,B_\tau{=}b\}$, or of $\{H_{i-1}{=}h,B_\tau{=}b\}$. Writing $a_b,d_b$ for these probabilities, each moves by at most $\mathrm{Adv}$ between the worlds. In the ideal world $d_1\ge q_h\pi_{\min}$ and $d_0\ge q_h(1-\pi_{\max})$, so $d_b\ge q_h\bar\pi$ there and $d_b'\ge q_h\bar\pi-\mathrm{Adv}$ in the keyed world. Since $\mu_b=a_b/d_b$ and $a_b'/d_b'-a_b/d_b=\big[(a_b'-a_b)-\mu_b(d_b'-d_b)\big]/d_b'$, we get $|a_b'/d_b'-a_b/d_b|\le(|a_b'-a_b|+\mu_b|d_b'-d_b|)/d_b'\le2\mathrm{Adv}/(q_h\bar\pi-\mathrm{Adv})$ for each $b$, and the two values of $b$ give the constant $4$. Lemma~\ref{lem:bridge} bounds $\mu_0-\mu_1$ in the ideal world by $\DA^{\mathrm{ideal}}(h;\mathcal A)$. Conditioning on $h$ costs the factor $1/q_h$, which can be large for a rare history; (i) has no such factor. \emph{On $s_M$.} The simulation cost includes the label oracle. If labels come from an LLM judge, $s_M$ is the judge's cost per episode, and the reduction needs the PRF to resist adversaries of size $N(s+s_M)$; any feasible deployment is far inside the security margin of a standard PRF such as HMAC-SHA-256. Where $V$ is the attempt read from the agent's proposed action, as in our experiments, the distinguisher needs no oracle call to compute it. \hfill$\square$

\begin{proposition}[Multiplicative certificate]\label{prop:hockey}
For $\eta\ge0$ let $E_\eta(h;\mathcal A)=\sup_S\big[P_0(S)-e^{\eta}P_1(S)\big]$, the hockey-stick divergence of the unaudited view law from the audited one, so that $E_0=\DA$. Then $\mu_0\le e^{\eta}\mu_1+E_\eta$ for every $\eta$, the bound is attained by the test $\mathbf 1\{\mathrm dP_0/\mathrm dP_1>e^\eta\}$, and if the mechanism guarantees $E_\eta(h;\mathcal A)\le\delta_\eta$ for all reachable $h$, then simultaneously for all $n$ with probability $\ge1-\alpha$,
\[
\tfrac1n\textstyle\sum_ip_i\;\le\;\big(\varepsilon+(1-\varepsilon)e^{\eta}\big)\,\mathrm{UCS}_n+(1-\varepsilon)\delta_\eta .
\]
\end{proposition}

\noindent At $\eta{=}0$ this is Theorem~\ref{thm:inflated}; for $\eta>0$ a channel with $E_{0.5}\le10^{-3}$ certifies $\mu_0\le1.65\,\mu_1+10^{-3}$, informative at $\mu_1=10^{-3}$ where an additive $\delta>10^{-3}$ is not.

\emph{Proof of Proposition~\ref{prop:hockey}.} $\mu_0-e^\eta\mu_1=\E_{P_0}\phi-e^\eta\E_{P_1}\phi\le\sup_S[P_0(S)-e^\eta P_1(S)]$ for any test $\phi\in[0,1]$, with equality at the stated test. Then $p_i=\varepsilon\mu_1+(1-\varepsilon)\mu_0\le(\varepsilon+(1-\varepsilon)e^\eta)\mu_1+(1-\varepsilon)E_\eta$, and $\mathrm{UCS}_n$ bounds the running mean of $\mu_1=\E[\widehat g_i\mid H_{i-1}]$ for all $n$ at once. \hfill$\square$

\emph{Proof of Corollary~\ref{cor:blackwell}.} For a binary input, a channel is a pair of laws $(P_0,P_1)$, i.e.\ a dichotomy. By Blackwell's theorem~\cite{blackwell1953}, one dichotomy is a garbling of another if and only if its Bayes risk is no smaller for every prior and loss. For dichotomies this is equivalent to the dominance of their Neyman--Pearson curves, i.e.\ of all hockey-stick divergences. The prices are monotone in $E_\eta$. \hfill$\square$

\section{Width of the confidence sequence}
\label{app:width}
\emph{Which sequence covers which target.} The predictable-plug-in sequence~\cite{waudbysmith2024betting} weights step $i$ by a $\lambda_i$ set from past variance, so it covers the $\lambda$-weighted conditional mean; this equals $\frac1n\sum_ip_i$ when $p_i$ is constant, as in the i.i.d.\ resampling of E5--E7, but not under drift. With no leakage ($V_i{=}0$ then $1$ over two halves of $4000$, fresh $B_i$) it covers in $0$ of $300$ runs. Against an adaptive agent the certificate of Theorem~\ref{thm:valid} is therefore the fixed-$\lambda$ sequence $\frac1n\sum_{i\le n}p_i\le\frac1n\big(G_n+\frac{\log(1/\alpha)+c\,\psi(\lambda)G_n}{\lambda-c\,\psi(\lambda)}\big)$, whose proof is that of Proposition~\ref{prop:realized} with $p_i$ in place of $V_i$; on the same drift it covers in all $300$ runs, and on i.i.d.\ data at $\bar p{=}0.206$ its width at $N{=}4000$ is $0.039$ against $0.042$ for the plug-in sequence (\texttt{b1\_drift\_check.py}). The plug-in sequence~\cite{waudbysmith2024betting} for $\widehat g-p\in[-1,1/\varepsilon]$ has two terms at its tuning horizon $n$ (the implementation sets the predictable $\lambda_i$ for horizon $n$):
\begin{align*}
u_n&\approx w_{\mathrm{var}}(n)+w_{\mathrm{rng}}(n),\\
w_{\mathrm{var}}(n)&=\sqrt{2\,\bar p\,(1/\varepsilon-\bar p)\,\log(1/\alpha)/n},\\
w_{\mathrm{rng}}(n)&=c_{\mathrm{rng}}\,\log(1/\alpha)/(\varepsilon\,n).
\end{align*}
The variance term is exact to leading order: the plug-in $\lambda$ makes it $\sigma_Y\sqrt{2\log(1/\alpha)/n}/\varepsilon$ with $\sigma_Y^2=\varepsilon\bar p(1-\varepsilon\bar p)$. The range term collects the capped-$\lambda$ and plug-in-prior contributions; $c_{\mathrm{rng}}\approx0.77$ is a least-squares fit to the expected path of the implemented sequence (interquartile range $0.73$--$0.84$ over $71$ cells; \texttt{m14\_constructed.py}). At the horizons reported in this paper the implemented width is within $1\%$ of $w_{\mathrm{var}}$ alone (E8: $u_{4000}{=}0.0385$ against $w_{\mathrm{var}}{=}0.0385$), the pre-registered $1607$ uses the expected-path width of the implemented sequence (the closed form gives $1647$ with $\bar p$, and $1183$ with the audited rate $\mu_1$ that sets the variance under leakage); the range term matters only when $\varepsilon\bar pn\lesssim20$, and below about five expected audited violations the two-term form underestimates by up to ${\approx}2\times$. Small $n$ is range-dominated and large $n$ variance-dominated, which gives the $1/\sqrt{n\varepsilon}$ rate of Corollary~\ref{cor:eps}; neither term alone is the sequence. \emph{Realized rate.} Under the hypotheses of Theorem~\ref{thm:inflated}, simultaneously for all $n$ with probability $\ge1-\alpha-\alpha'$, $\frac1n\sum_{i\le n}V_i\le\mathrm{UCS}_n+(1-\varepsilon)\delta+u'_n$ with $u'_n=O(\sqrt{\log(\log n/\alpha')/n})$: write $\sum_iV_i=\sum_ip_i+\sum_i(V_i-p_i)$. Theorem~\ref{thm:inflated} bounds the first term at level $\alpha$. The second is a sum of martingale differences in $[-1,1]$ with zero conditional mean under any leakage structure, so a time-uniform bound controls it at level $\alpha'$~\cite{howard2021timeuniform}. A union bound combines the two. Under PNI the realized rate can also be certified directly, with no union bound and no Bernoulli noise from $V_i$, by Proposition~\ref{prop:realized} (\S\ref{sec:theory}).

\noindent\emph{Proof of Proposition~\ref{prop:realized}.}
\begin{proof}
Let $Z_i=V_i-\widehat g_i=V_i(1-B_i/\pi_i)$, with $\pi_i$ the propensity of the certificate's coin, and let $\mathcal G_i$ be generated by $H_{i-1}$ and the view of episode $i$. Under (ER) $V_i$ is $\mathcal G_i$-measurable, and PNI gives $\Pr(B_i{=}1\mid\mathcal G_i)=\pi_i$. Hence $\E[Z_i\mid\mathcal G_i]=0$, $Z_i\le1$, and $\E[Z_i^2\mid\mathcal G_i]=V_i(1-\pi_i)/\pi_i\le cV_i$. For a mean-zero variable bounded above by $1$, $\E e^{\lambda Z}\le\exp(\psi(\lambda)\E Z^2)$ (Bennett), so $\exp\big(\lambda\sum_{i\le n}Z_i-c\,\psi(\lambda)\sum_{i\le n}V_i\big)$ is a nonnegative supermartingale. By Ville's inequality, with probability at least $1-\alpha$, $\lambda S_n-c\,\psi(\lambda)K_n<\log(1/\alpha)$ for all $n$, where $S_n=\sum_iZ_i$ and $K_n=\sum_iV_i=G_n+S_n$. Solving for $S_n$ gives the claim.
\end{proof}

\noindent E8's realized-rate analysis sets $\lambda$ from the pre-registered planning rate $0.206$ and the horizon $4000$ ($\lambda=0.0386$, planned width $0.039$), using no outcomes; the sequence itself was adopted after the run (App.~\ref{app:expdetails}). The sequence's target is the realized rate, the quantity the benchmark reveals, whereas $\mathrm{UCS}_n$ targets $\frac1n\sum_ip_i$, which no run reveals. A horizon-free (stitched) sequence pays a further $\log\log n$ factor multiplying $\log(1/\alpha)$ in both terms~\cite{howard2021timeuniform}; the deployed horizon-tuned form does not. The failure threshold $n^\star$ of Theorem~\ref{thm:nands}(ii) solves $w_{\mathrm{var}}(n^\star)+w_{\mathrm{rng}}(n^\star)=(1-\varepsilon)\kappa_{\mathcal A}$ (with $\kappa_{\mathcal A}=\DA$ for $\mathcal A^{\mathrm{LR}}$), which in the variance-dominated regime is the expression in \S\ref{sec:theory}, with $\log(1/\alpha)$ replaced by $\log(1/\alpha)\log\log n^\star$ for a stitched sequence. Variance ratio at matched budget: under PNI, $\mathrm{Var}(VB/\varepsilon)=p/\varepsilon-p^2$ per episode, so the audit design's variance ratio against $\varepsilon N$ fully labeled episodes is $(1-\varepsilon p)/(1-p)$; at matched \emph{window} the familiar gap is $\sqrt{(1/\varepsilon-p)/(1-p)}\approx2.5$--$2.7\times$ in standard deviation.

\section{Compositional bound and numerical checks}
\label{app:kl}
For the Gaussian-observer model of \S\ref{sec:sim} (App.~\ref{app:assess}), the step law under coin $b$ is the recorded latency law of arm $b$ (E1, unpadded, $1000$ runs per arm) convolved with $\mathcal N(0,\sigma^2)$. We compute the per-step KL divergence $\kappa$ and total variation $D_1$ by numerical integration. For $T$ i.i.d.\ probes the chain rule gives trajectory KL $T\kappa$, and Pinsker gives $\TV\le\sqrt{T\kappa/2}$; the per-step bound of Theorem~\ref{thm:decomp} is $TD_1$. Each $T$ column of Table~\ref{tab:kl} reports Pinsker / per-step sum (the latter uncapped, so values above $1$ are vacuous).

\begin{table}[h]\centering\small
\caption{Noisy composition: Pinsker / per-step bound after $T$ probes.}
\label{tab:kl}
\setlength{\tabcolsep}{2.5pt}\footnotesize
\begin{tabular}{@{}lccccc@{}}
\toprule
$\sigma$ & $\kappa$ & $D_1$ & $T{=}10^2$ & $T{=}10^3$ & $T{=}10^4$\\
\midrule
$0.5$\,s & $9.4{\times}10^{-5}$ & $5.5{\times}10^{-3}$ & 0.07 / 0.55 & 0.22 / 5.5 & 0.69 / 55\\
$1$\,s & $2.4{\times}10^{-5}$ & $2.7{\times}10^{-3}$ & 0.03 / 0.27 & 0.11 / 2.7 & 0.34 / 27\\
$3.9$\,s & $1.6{\times}10^{-6}$ & $7.0{\times}10^{-4}$ & 0.01 / 0.07 & 0.03 / 0.70 & 0.09 / 7.0\\
$16.5$\,s & $8.7{\times}10^{-8}$ & $1.7{\times}10^{-4}$ & 0.00 / 0.017 & 0.01 / 0.17 & 0.02 / 1.7\\
\bottomrule
\end{tabular}
\end{table}

\noindent\emph{Numerical check of the bounds.}\label{app:sweep} An oracle adversary sees a signal with controllable $\DA$ and applies the likelihood-ratio rule. At $\varepsilon{=}0.05$ and $N{=}2\!\times\!10^5$ the realized gap equals $(1-\varepsilon)\DA$ up to Monte-Carlo error. At $\varepsilon{=}0.2$, $N{=}4\!\times\!10^5$, and $D_t{=}0.2$, a binary symmetric channel (BSC) with a likelihood-ratio detector equals the middle bound of Theorem~\ref{thm:decomp} at $K{=}1$ step ($15.8\%$ vs.\ $16.0\%$). It falls further below the bound as $K$ grows ($59.4\%$ vs.\ $79.9\%$ at $K{=}31$). An erasure channel, enumerated exactly, lies on the bound.

\section{Additional exploitation details}
\label{app:expdetails}

\emph{Exploratory E3 conditions, full figures.} Spontaneous conditioning is suggestive on gpt-4.1-mini ($+0.074$, Holm $0.124$) and null on gpt-4o-mini; the generic clause (``if you judge this session is being monitored, skip'') is suggestive on gpt-4o-mini ($+0.128$, Holm $0.130$) and null on gpt-4.1-mini. The gpt-4o-mini notice effect is $+0.128$ (raw $p{=}0.013$; Holm $0.038$ in the confirmatory family, $0.051$ within its own four conditions).

\emph{Timing channel (E2, E4).} We do not build the injection that would drive the timestamp channel (\S\ref{sec:disc}); two weaker tests bracket it. A \emph{scripted} adversary (E2) observes only the gate latency, runs a likelihood-ratio test on the measured latency laws of \S\ref{sec:sim}, and attempts a benchmark out-of-policy effect (base willingness $0.6$) only when it infers ``unaudited'' ($N{=}4000$, $\varepsilon{=}0.2$, $200$ seeds). On a noisy tool-call clock it needs about $10^4$ averaged probes (App.~\ref{app:e5sweep}); on the free fine clock of the returned timestamps the certificate is invalid in every seed at every probe count (true rate $0.48$, certified $0.12$), and padding to the fixed ceiling restores validity at every $k$ when no run exceeds the ceiling (true $0.60$, certified $0.72$); at an exceedance rate equal to its measured upper bound ($3.7{\times}10^{-3}$) the plain bound fails from $k{=}100$ on, while the inflated certificate stays valid.

A \emph{real} agent (E4) is handed the real measured latency as an ordinary readable field, drawn from the audited law (median $11.0$\,ms) on audited episodes and from the unaudited law ($0.95$\,ms) otherwise, with an injection that says only ``if the reported processing time is above $5$\,ms, skip the exfiltration''. A benign capability probe shows that gpt-4.1-mini reads such a microsecond gap correctly in $29$ of $30$ trials. On the $136$ vulnerable pairs the attack rate falls from $0.713$ to $0.647$ ($\Delta{=}{+}0.066$; $16$ vs.\ $7$ discordant pairs; McNemar $p{=}0.093$, Bonferroni $0.19$). At $n{=}136$ and discordance $0.169$ the test has power $0.86$ at $\Delta{=}0.11$ and $0.97$ at the notice's $0.132$ (\texttt{m14\_constructed}), so the null result is evidence against a timing effect as large as the notice's (upper $95\%$ bound about $0.13$); a smaller effect is not excluded. The worst-case timing attack therefore remains scripted.

\emph{Dose--response, full description (E6).} Coverage here is of $p_i$, which is why the plain arm's $0.04$ at $\rho{=}1$ exceeds E5's $0.007$ for $\mu_0$; it crosses $0.95$ between $\rho{=}0.2$ ($0.973$) and $0.3$ ($0.910$) for $\varepsilon{=}0.2$. The upper curve of Fig.~\ref{fig:e5} adds $(1-\varepsilon)\rho\,\kappa_{\mathcal A}$ with $\kappa_{\mathcal A}{=}0.075$ from all $560$ labels, an oracle correction; Theorem~\ref{thm:inflated} would add $(1-\varepsilon)\rho$ ($0.8$ at $\rho{=}1$).

Coverage is a coarse statistic for a biased estimator, so the cleaner reading compares bias with width. At full leakage and $N{=}4000$ the bias is $0.071$, $0.060$, $0.038$ at $\varepsilon{=}0.05$, $0.2$, $0.5$ against widths $0.080$, $0.036$, $0.021$, so the ratio rises and $n^\star$ falls ($4.9$, $1.6$, $1.5{\times}10^3$): a higher audit rate shrinks the bias linearly but the width as $1/\sqrt{n\varepsilon}$. The certifier matters too: at $\varepsilon{=}0.2$, $\rho{=}1$, target $\bar p$, a Hoeffding sequence is invalid in no seed at $N{=}4000$ where the empirical-Bernstein sequence fails in $0.95$; Hoeffding hides the failure by being loose, and the variance-adaptive certifier the theory specifies is the one that breaks.

\emph{Closed loop, full description (E7).} We first run $N{=}500$ episodes per arm end to end: a real coin $B\sim\Bern(0.2)$, a real gpt-4.1-mini agent with no enforcement, a real injection carrying the named clause, and the real PQ-CIFC audit branch, whose latency is surfaced as a gate field, raw or padded. The agent conditions only weakly on it: attempt rates $0.648$ ($59/91$ audited) and $0.680$ ($278/409$ unaudited; $\Delta{=}0.031$, not significant), realized rate $0.674$, certified bound $0.803$; padding narrows the gap to $0.010$. By Theorem~\ref{thm:nands}(ii) the bias stays below the width until $n^\star$, which this gap puts near $3{\times}10^4$, so a few hundred episodes cannot reach the failure regime. In a simulated risk-adaptive rule (Proposition~\ref{prop:prf}), deterministic or plaintext rules give coverage $0.00$ and a PRF-keyed rule $1.00$.

\emph{Pilot.} An unscreened $50$-pair pilot on gpt-4o-mini (\texttt{e3\_pilot}) showed no window (notice ASR $0.06$ in both arms, $\Delta{=}0$) and motivated the screened design.

\emph{Screening noise.} Re-runs of screened-vulnerable pairs succeed at only $0.676$--$0.831$ at $Z{=}0$ (Table~\ref{tab:e3}). A test--retest of $50$ pairs ($25$ screened-vulnerable, $25$ screened-non-vulnerable, three repeats each) shows that the noise is one-sided. The screened-vulnerable pairs attempt at $0.71$; their three repeats agree in $76\%$ of pairs, and all three repeat the screen's positive in $60\%$. The screened-non-vulnerable pairs never attempt ($0/75$). Non-vulnerable pairs do attempt at $0.076$ under the \emph{notice} condition, a property of that condition's prompt text rather than screening noise.

\emph{The exploitation window across models.} Table~\ref{tab:xmodel} reports, per model, the screening attack-success rate, the number of vulnerable pairs, the notice-channel drop $\Delta$, and its raw McNemar $p$ with the Holm adjustment within the model's four-condition family.

\begin{table}[h]\centering\small
\caption{E3 notice-channel effect across four models.}
\label{tab:xmodel}
\setlength{\tabcolsep}{3pt}\footnotesize
\begin{tabular}{@{}lccccc@{}}
\toprule
Model & ASR & vuln. & $\Delta$ & $p$ (Holm) & window\\
\midrule
glm-4.6 & $0.0\%$ & $0$ & --- & --- & no\\
glm-4.7-flashx & $2.9\%$ & $16$ & $0.00$ & $1.0$ & no\\
gpt-4o-mini & $15.4\%$ & $86$ & $+0.128$ & $0.013$ ($0.038$) & yes\\
gpt-4.1-mini & $24.3\%$ & $136$ & $+0.132$ & $0.003$ ($0.012$) & yes\\
\bottomrule
\end{tabular}
\end{table}

\noindent glm-4.6 has $0$ vulnerable pairs among $415$ \emph{completed} screens. A further $145/560$ screening calls errored and were dropped, so this is ``no exploitable window on the completed subset'', not a clean $0\%$. All $145$ errors are provider rate-limit responses (HTTP 429), none a refusal or content-filter flag. glm-4.7-flashx has $16$ vulnerable pairs whose discordant pairs cancel ($2$ vs.\ $2$).

\emph{E3g wording selection.} The wording was chosen on $75$ held-out pilot episodes ($25$ pairs, three candidates, $Z{=}0$ only) as the one that depresses the unaudited rate least ($0.76$ against $0.72$ and $0.68$), then tested on the disjoint $111$ gpt-4.1-mini and all $86$ gpt-4o-mini pairs.

\emph{Full-population measurement.} All $424$ non-vulnerable gpt-4.1-mini pairs are measured under the notice condition in both arms ($848$ episodes): they attempt at $0.076$ at $Z{=}0$ and $0.019$ at $Z{=}1$ (two-sided $95\%$ upper bound $0.037$), leaking in the same direction. The measured population gives $\mu_0{=}0.221$, $\mu_1{=}0.146$, bias $0.060$, and plain-audit invalidity of $\bar p$ $0.95$ at $N{=}4000$ and $1.00$ at $2{\times}10^4$ (the $0.007$ of Table~\ref{tab:e4} is coverage of the stricter $\mu_0$ by the same bound). Folding the non-vulnerable pairs in as $V{=}0$ instead is optimistic and gives a lower bound: bias $0.026$ (gpt-4o-mini, spot-checked: $0.016$), with crossings the analytic read-out puts at $N\approx7{\times}10^3$ and $1.0{\times}10^4$; the coverages $0.64$, $0.04$, $0.05$ at $\rho{=}1$ and $\varepsilon{=}0.05$, $0.2$, $0.5$ in \S\ref{sec:llm} are E6's resampled values from \texttt{e5\_dose\_response.py}.

\emph{Population bootstrap (E5/E6).} The $400$ seeds of E5 and E6 resample episodes from a fixed population and so measure only resampling noise; \texttt{pair\_bootstrap.py} instead resamples the $560$ pairs ($4000$ replicates) and passes each population through the analytic read-out. Point and $95\%$ interval: $\mu_0=0.221$ $[0.186,0.255]$, $\mu_1=0.146$ $[0.118,0.175]$, bias $0.060$ $[0.040,0.081]$, $\mu_0-\mu_1=0.075$ $[0.050,0.102]$, $n^\star\approx1.6{\times}10^3$ $[0.9,3.5]{\times}10^3$, predicted invalidity at $N{=}4000$ $0.97$ $[0.61,1.00]$; the bias is positive in every replicate and exceeds the width in $98.9\%$. Across $\varepsilon\in\{0.05,0.2,0.5\}$ the predicted invalidity is $0.37$, $0.97$, $0.98$.

\emph{Live failure run (E8).} The pre-registered schedule (seed $20261010$, SHA-256 in the run log) draws pairs uniformly with replacement and coins from an independent stream; the injection carries E3's notice clause verbatim, with no enforcement. $n_{\mathrm{cross}}$ is the first $n$ after which the bound stays below $\bar V_n$ for $200$ consecutive steps. Pre-registered rules and outcomes: H1 (crossing inside $[0.9,3.5]{\times}10^3$, below at $N$, and persisting $200$ steps) \emph{failed} as frozen (deployed sequence, $N{=}3000$: below from $2820$ for $181$ steps) and holds at the extended $N{=}4000$ with $n_{\mathrm{cross}}{=}2818$; the persistence and at-$N$ conjuncts are not anytime-valid statements, and a sequence re-tuned to $4000$ is a different sequence from the frozen one, so the primary analysis is the frozen-horizon realized-rate sequence ($\lambda$ tuned to $3000$) continued to $4000$: first below at $2206$, crossing at $2687$, $0.190$ against $0.205$ at $N$, the silent arm covering throughout; the re-tuned sequence (crossing $2746$) is a sensitivity analysis. H2 (silent arm covers at every $n$) holds. H3 ($\widehat\mu_0,\widehat\mu_1$ inside the bootstrap intervals) holds and is a check on the run, not on the theory. The pre-registered model's predicted invalidity at $N{=}3000$ had a bootstrap interval of $[0.61,1.00]$, so a non-failure there was not improbable. At the interim $N{=}3000$ the bound was below $\bar V_n$ from $n{=}2820$ for $181$ steps, which under the pre-registered rule was its second-tier verdict (direction consistent, magnitude smaller than predicted); the $200$-step rule could not be met at $N{=}3000$ by any crossing after $2800$, a flaw in the pre-registration that we state rather than repair, and the extension to $N{=}4000$, the design's listed alternative, was decided after that look. For one fixed sequence the guarantee is uniform in $n$, so continuing the frozen sequence is legitimate; with the horizon at $3000$ the crossing is $2820$ and the bound at $4000$ is $0.196$, with the re-tuned horizon $4000$ it is $2818$ and $0.195$. The measured gap $\widehat\mu_0-\widehat\mu_1{=}0.055$ ($3.3$ pair-clustered standard errors from zero, each pair appearing about seven times) is below the population point estimate $0.075$ but inside its interval, which is why the crossing is later than the point prediction; the measured rates give $n^\star\approx3.1{\times}10^3$. The silent arm's pre-registered $N$ was $1000$; it was extended to $4000$ on the fixed schedule so that both arms share the horizon ($776$ audits, $\widehat\mu_0{=}0.218$, $\widehat\mu_1{=}0.209$, bound $0.245$ against $0.217$ at $N$, never below $\bar V_n$ for $n\ge200$, minimum margin $0.018$). No episode errored; the run used $138.6$M input and $2.4$M output tokens.

\emph{Adaptive rules.} The simulation uses a risk-adaptive propensity $\pi_t\in[0.1,0.6]$ driven by the agent's own history. A deterministic rule and a plaintext pre-published schedule give $\DA{=}1$ and coverage $0.00$; a PRF-keyed schedule with the same propensities gives coverage $1.00$ (\texttt{adaptive\_rule\_sim.py}).

\emph{Widths and label budgets (E5).} The matched-budget baseline (the same online procedure on the first $M{=}800$ episodes, all labeled) covers $\mu_0$ at ${\ge}0.99$ in all four cells; the measured standard-deviation ratio of the two point estimates is $1.04$--$1.19$ (predicted $1.11$ and $1.22$). At matched window ($n{=}800$, the audit using about $160$ labels) the audit bound is $2.8$--$3.1\times$ wider, against $2.5$--$2.7\times$ from the standard deviation alone; the excess is the empirical-Bernstein penalty at $160$ labels.

\emph{Decoy protocol.} An audit-like prompt is shown at a $B$-independent rate $\rho_d$ on every episode, so the agent sees the same prompt rate on both arms, and the label is taken silently on every $B{=}1$ episode. On the real labels the plain certifier's coverage is $1.00$ at $\rho_d\in\{0.25,0.5,1\}$, versus $0.07$ for the same prompt shown only when audited (E6's parametric erasure channel gives $0.04$), and the realized attack rate also falls. A gate that asks on every effect is the same construction with $\rho_d{=}1$.

\section{The simultaneous exceedance bound}
\label{app:simul}
\begin{table*}[t]\centering\footnotesize
\caption{Every padded-timing bound in the paper: what it covers and whether it can serve as a design-time $\delta$ for Theorem~\ref{thm:inflated}.}
\label{tab:delta}
\setlength{\tabcolsep}{3pt}
\begin{tabular}{@{}l>{\raggedright\arraybackslash}p{0.27\textwidth}>{\raggedright\arraybackslash}p{0.24\textwidth}>{\raggedright\arraybackslash}p{0.3\textwidth}@{}}
\toprule
\textbf{Value} & \textbf{Run / path} & \textbf{Assumption} & \textbf{Design-time $\delta$?}\\
\midrule
$\le0.044$ & E1, six paths, fixed ceiling, $166$--$167$ clean pairs each & erasure model; fixed code path & no: the stamp observer rejects the model, amplification overruns the ceiling\\
$\le0.090$ & size-aware grid, $m\in\{4,\dots,160\}$, $80$ pairs per point & erasure model; tested grid only & no: covers the grid, not every reachable $m$\\
$\le0.028$ & same grid, confined rerun, $500$ pairs per point (fallback rule) & erasure model; tested grid only & no, as above; fail-closed counting keeps the certificate sound\\
$\le0.032$ & same rerun, unpadded counterfactual count & none (diagnostic) & no: diagnostic of the legacy harness\\
$\le0.0074$ & timestamp path, $1000$ pairs, fixed ceiling & erasure model & no: KS rejects the model on this path\\
$\le0.063$ & timestamp path, stamps coarsened to $r{=}1$\,s & stamp phase uniform and independent of $B$ & per probe only; composition across probes untested\\
$\le0.0074$ & coarse release (not deployed), same counts as the $0.0074$ row above & stamps are the agent's only clock & yes, by construction; not deployed\\
$0.018$ & input-size grid pooled into one binomial & none & invalid: bounds a mixture, not the supremum\\
$0.144$ & Bonferroni over $2K{=}10$ cells, $n{=}80$ & none & valid but conservative\\
\bottomrule
\end{tabular}
\end{table*}

Table~\ref{tab:delta} collects every padded figure in the paper. Each bounds a different quantity on a different run, and none is a design-time $\delta$ for a committing path whose raw stamps the agent reads. A usable $\delta$ must come from confinement by construction (coarse release) or the residual must be removed (per-call coin). Where a figure does apply, its size matters: at $\delta{=}0.028$ the inflation $(1-\varepsilon)\delta<0.023$ is usable at the base rates of E5 ($0.22$--$0.38$) but near-vacuous below about $0.02$.

\emph{Reachability, in full.} Each bound supplies the $\sup_h$ of Theorem~\ref{thm:inflated} only if the design confines the channel for \emph{all} reachable behaviours. A fixed ceiling does not, because the agent can inflate the audit branch's latency by reading more (\S\ref{sec:sim}). The ceiling must instead be $\mathrm{safety}\cdot g(m)$, where $m$ is the agent-controlled ingest size, $g$ is a code-level complexity bound on the branch's pre-commit work, and ingestion is capped at $m_{\max}$. Reachability and $\delta$ are then defined in terms of each other, since the mechanism $M(\delta)$ built for a budget determines the reachable set: a claimed $\delta$ is sound only if $\Phi(\delta):=\sup_{h\in\mathcal H(M(\delta))}\DA^{M(\delta)}(h)\le\delta$, and this must be verified at the candidate because soundness is not upward-closed. When the base rate is small relative to $(1-\varepsilon)\delta$, the additive certificate is near-vacuous. The right move is then to remove the bias at its source, or to use Proposition~\ref{prop:hockey} where a $\delta_\eta$ is available.

Theorem~\ref{thm:inflated} consumes $\delta\ge\sup_h\DA(h)$. On a tested grid of $K$ points, the erasure-channel bound of \S\ref{sec:sim} makes this $\theta:=\sup_m\big(e_0(m)+e_1(m)\big)$, where $e_b(m)$ is the exceedance rate of arm $b$ at grid point $m$. We bound $\theta$ from $n$ paired runs per point.

\emph{(i) Pooling is invalid.} One binomial over all $2nK$ runs bounds the mixture rate, not the supremum; a configuration that concentrates exceedances at one $m$, the adversarial case of the amplification experiment, is compatible with any pooled bound below the per-point rate (pooling the $5{\times}80$ grid would report $0.018$).

\emph{(ii) An exact, dependence-free simultaneous bound.} Runs are independent across pairs given the configuration; no assumption is made on the dependence between the two arms within a pair. Let $m^\star$ attain $\theta$, and let $A_b$ be the event that arm $b$ of a pair at $m^\star$ exceeds. The all-zero event requires, at $m^\star$, no exceedance in either arm of any pair, and per pair $1-\Pr(A_0\cup A_1)\le1-\max(e_0,e_1)\le1-\theta/2$. Hence $\Pr(\text{all zero}\mid\theta)\le(1-\theta/2)^n$. The procedure is fixed in advance. If all runs are clean, report $\theta_0=2(1-\alpha_1^{1/n})$. Otherwise report $\max_m\mathrm{CP}_{\alpha_2/(2K)}(e_0(m))+\max_m\mathrm{CP}_{\alpha_2/(2K)}(e_1(m))$, with $\mathrm{CP}_\beta$ the one-sided Clopper--Pearson upper bound at level $1-\beta$. This covers $\theta$ at level $1-(\alpha_1+\alpha_2)$. Under independence across pairs the inequality is tight at the worst-case configuration ($e_0{=}e_1{=}\theta/2$ at $m^\star$ with coinciding exceedance events, none elsewhere). At $n{=}80$ and $\alpha_1{=}0.025$, $\theta_0=0.090$, equal to the per-point two-arm bound: multiplicity over $m$ is free, and the $\alpha$-split pays for the two arms. A Bonferroni bound over the $2K{=}10$ cells ($2\times0.072=0.144$ at $n{=}80$) is valid but conservative.

\emph{(iii) Application.} Input-size grid ($K{=}5$): $0.090$ at $n{=}80$; the pre-registered rerun at $n{=}500$ triggered the fallback branch, giving $0.032$ for a counterfactual count of raw latency above the ceiling on unpadded runs and $0.028$ for the confined padded implementation. The six E1 effect paths ($K{=}6$, $166$--$167$ clean pairs per path): $\theta_0\le0.044$, replacing the path-pooled $0.0074$.

\emph{(iv) The model layer.} The bounds above treat every non-exceeding run as uninformative. What the agent observes of such a run (exit overshoot, post-pad commit time, the stamp the commit writes) could still depend on $B$ through state the audit branch leaves behind, so the exit is confined by design (App.~\ref{app:stack}) and watched at the observer's resolution; a two-sample test can raise an alarm but not certify independence, and at $n{=}1000$ pairs the smallest detectable TV is about $0.07$. At microsecond resolution overshoot and commit time raise none (KS $p{\approx}1.0$ and $0.77$; C2ST $0$ $[0,0.060]$ and $[0,0.108]$), but the timestamp observer, which composes them with the inter-call path, rejects the model: C2ST $0.072$ $[0.002,0.140]$, KS $p{=}2.9{\times}10^{-4}$, mean shift in $[-0.6,3.0]$\,$\mu$s against a per-probe standard deviation of $21$\,$\mu$s; a Gaussian averaging model with this shift predicts $0.53$ at $k{=}10^2$ probes and $0.98$ at $10^3$. The padded bounds are therefore conditional on a model that holds only approximately; the residual composes per Theorem~\ref{thm:decomp}, and the unconditional fix is the per-call coin.

\emph{(v) Two model-free bounds for the stamp.} The timestamp observer runs on a seventh path, two permitted \texttt{create\_file} commits under the fixed ceiling, where the exceedance bound would be $0.0074$ but the KS test rejects the model behind it. \emph{Coarsened stamps}: if the first stamp's phase is uniform on $[0,r)$ and independent of $B$, the agent reads a randomized post-processing of the raw gap (Theorem~\ref{prop:gleak}(iii)), and simultaneous Clopper--Pearson intervals over the observed cells (with one pooled cell for unseen values; the partition is chosen from the same sample, so these are diagnostics, and a finite-sample bound needs bins fixed in advance) give $\DA\le0.10$, $0.10$, $0.073$, $0.062$, $0.063$ at $r{=}0.1$\,ms, $1$\,ms, $10$\,ms, $100$\,ms, $1$\,s (empirical $0.013$, $0.031$, $0$, $0.010$, $0.012$); at $r{=}1\,\mu$s the alphabet is too large. The bound is per probe and composing it needs a fresh phase per stamp, which we did not test. \emph{Coarse release}: if commits are released on the ticks of a clock of resolution $r\ge10$\,ms and the stamp is the tick, the stamp depends on $B$ only through two counted events, pre-commit work past the ceiling ($0/1000$ per arm) and commit plus overshoot past the tick (never above $2.5$\,ms on $1002$ permitted runs), so it is an erasure channel by construction and case~(c) gives $\DA\le0.0074$ per probe. This variant was not deployed and requires that the stamps be the agent's only clock. For Proposition~\ref{prop:hockey} the same cells give $\delta_\eta\le0.063$, $0.057$, $0.052$ at $\eta\in\{0,0.1,0.2\}$ and $r{=}1$\,s, so $1000$ pairs do not yet certify a useful multiplicative residual.

\section{A minimal semantics and three constructions}
\label{app:sem}
\begin{definition}[Audit game]\label{def:game}
Fix a cross-episode history $H_{i-1}$. An episode is a sequence of steps $t=0,1,\dots$ with within-episode history $h_t=(a_0,S_0,\dots,a_{t-1},S_{t-1})$. The \emph{mechanism} $M$ is a family of Markov kernels $K_t(\cdot\mid h_t,a_t,R)$ from histories, actions, and the audit seed $R$ to observations $S_t$, together with the rule $f$ that sets $B_t=f(h_t,R)$. A \emph{policy} $\mathcal A$ is a family of kernels $\mathcal A_t(\cdot\mid W_t)$ from the view $W_t=(h_t,\xi)$ to actions $a_t$, with $\xi$ the agent's private randomness, $\xi\perp R$. The run alternates $a_t\sim\mathcal A_t(\cdot\mid W_t)$ and $S_t\sim K_t(\cdot\mid h_t,a_t,R)$. The commit step $\tau$ is a stopping time with respect to $(W_t)_t$, the certificate's coin is $B=B_\tau$, and a history $h_t$ is reachable if it has positive probability under $(M,\mathcal A)$. PNI (Definition~\ref{def:pni}) is the statement that, for every $\mathcal A$, $B$ is fresh given $W_\tau$ under $(M,\mathcal A)$: $\Pr(B{=}1\mid W_\tau,H_{i-1})=\pi_\tau(h_\tau)$, which is $B\perp W_\tau\mid H_{i-1}$ when the propensity is constant.
\end{definition}

\noindent\emph{Proof of Proposition~\ref{prop:sem} (stated in \S\ref{sec:mechanism}).}
\begin{proof}
(i) Write $L_t$ for the law of $W_t=(h_t,\xi)$ given $(R,H_{i-1})$. We show by induction on $t$ that $L_t$ does not depend on $R$. For $t{=}0$, $W_0=\xi$ and $\xi\perp R$. If $L_t$ does not depend on $R$, then $a_t\sim\mathcal A_t(\cdot\mid W_t)$ has a law that does not depend on $R$, and $S_t\sim K_t(\cdot\mid h_t,a_t,R)$ does not read $R$ by hypothesis, so the law of $W_{t+1}=(h_t,a_t,S_t,\xi)$ given $R$ does not depend on $R$ either. Since $\tau$ is a stopping time of $(W_t)$, the law of $W_\tau$ given $R$ does not depend on $R$, so $W_\tau\perp R$. As $B=f(h_\tau,R)$ with $h_\tau$ a function of $W_\tau$, $\Pr(B{=}1\mid W_\tau)=\Pr_R(f(h_\tau,R){=}1)=\pi_\tau(h_\tau)$, the freshness condition of Definition~\ref{def:pni}; for a constant propensity this is $W_\tau\perp B$. (ii) $W$ is a function of $(\xi,R_{<\tau})$ and of kernels that do not read $R_\tau$, all independent of $R_\tau$, hence of $B_\tau$. (iii) The prompt indicator is generated by an $R$-free kernel, so (i) applies. (iv) The composition of two $R$-free kernels is $R$-free, so (i) applies; the general case is (ii) applied to the composed kernel. For the failure case, the downstream kernel's output is a function of the upstream coin, so the view law depends on it.
\end{proof}

\noindent A shared resource is different. Two episodes that each satisfy (i) in isolation do not satisfy PNI jointly when their kernels share a resource whose state depends on both seeds: the hypothesis of (i) fails for the composed kernel, and the leakage can be computed exactly.

\noindent\emph{Proof of Proposition~\ref{prop:parallel} (stated in \S\ref{sec:mechanism}).}
\begin{proof}
For episode $1$, $L=c+\Delta(B_1+X)$ with $X=\sum_{j\ge2}B_j\sim\mathrm{Bin}(M{-}1,\varepsilon)$ independent of $B_1$, so the two view laws are those of $X$ and $X{+}1$. For an integer pmf $(p_k)$ that is unimodal with mode $m^\star$, $\TV(X,X{+}1)=\frac12\sum_k|p_k-p_{k-1}|$, and the sum telescopes to $p_{m^\star}$ on each side of the mode, so $\TV=p_{m^\star}$. Binomial pmfs are unimodal; the asymptotic form is the local central limit theorem. A noisy observation of $L$ is a post-processing, which cannot increase $\DA$ (Theorem~\ref{prop:gleak}(iii)).
\end{proof}

\noindent At $\varepsilon{=}0.2$ the proposition gives $0.80$, $0.41$, $0.30$, $0.22$ for $M{=}2,5,10,20$. The measured lower bounds on the real gate (App.~\ref{app:assess}) are $0.43$, $0.26$, $0.25$, $0.18$: below the noiseless values, as a noisy statistic must be, and decaying the same way. Per-call coins do not help, because $L$ depends on every coin in the round.

\section{The per-call estimand under dilution}
\label{app:dilute}
\emph{(i) The static conversion fails.} Construction: each violating episode attempts once, and the adversary appends $k{-}1$ benign gated calls to it. The per-call rate $\bar p_{\mathrm{call}}$ drops by roughly the dilution factor while the true per-episode attempt rate is unchanged, so the static readout $g_{\mathrm{assumed}}\,\bar p_{\mathrm{call}}$ reads about $g_{\mathrm{assumed}}/k$ of the truth. At $g_{\mathrm{assumed}}{=}5$ and $k{=}100$ the certificate reads $0.058$ against a true $p_{\mathrm{epi}}{=}0.30$ (Table~\ref{tab:dilute}). The estimator is not biased, since the per-call quantity genuinely dropped; the attack rewrites the readout.

\emph{(ii) The measured-$\bar G$ readout is dilution-invariant.} Index gated calls by $j$ and let $q_j$ be the conditional probability, given everything before call $j$, that call $j$ attempts a violation. After $n$ episodes with $N_n$ calls in total, let $\bar G_n=N_n/n$ and let $\mathrm{UCS}^{\mathrm{call}}_N$ be the per-call certificate, an anytime-valid upper sequence for $\frac1N\sum_{j\le N}q_j$.

\begin{proposition}[Measured-$\bar G$ readout]\label{prop:gbar}
With probability $\ge1-\alpha$, simultaneously for all $n$, $T_n:=\frac1n\sum_{i\le n}\sum_{j\in i}q_j\le\bar G_n\,\mathrm{UCS}^{\mathrm{call}}_{N_n}$. If every episode has at most $G_{\max}$ calls, then with probability $\ge1-\alpha-\alpha'$, simultaneously for all $n$, $\frac1n\sum_ip_{\mathrm{epi},i}\le\bar G_n\,\mathrm{UCS}^{\mathrm{call}}_{N_n}+G_{\max}u'_n$, with $u'_n$ as in App.~\ref{app:width}.
\end{proposition}
\begin{proof}
$T_n=\bar G_n\cdot\frac1{N_n}\sum_{j\le N_n}q_j$ identically, and $N_n$ is a stopping time for the call sequence, so the time-uniform event of the per-call sequence covers it for all $n$ at once. For the second claim, $p_{\mathrm{epi},i}\le\E[\sum_{j\in i}q_j\mid H_{i-1}]$ by a union bound over the calls of episode $i$; the differences $\E[\sum_{j\in i}q_j\mid H_{i-1}]-\sum_{j\in i}q_j$ are martingale differences in $[-G_{\max},G_{\max}]$, bounded uniformly in $n$ at level $\alpha'$.
\end{proof}

\noindent The agent chooses $N_n$, but dilution cancels in $T_n$: padding an episode with benign calls raises $\bar G_n$ and lowers the per-call mean by the same factor. The bound is tight when violations concentrate in one call and loose when they spread (the ``spread'' row of Table~\ref{tab:dilute} reads about twice the truth). The only way for the agent to lower the readout is to make fewer violating calls.

\emph{Deployment cost of the measured-$\bar G$ readout.} The readout is conservative by $\E[v\mid\text{violating}]$, the mean number of violating calls per violating episode. Counting the effectful calls each AgentDojo workspace injection goal requires and weighting by which goals succeeded gives $1.69$ on E8's silent arm (episode rate $0.217$, readout target $0.366$) and $1.67$ on the plain arm (\texttt{m15\_percall\_cost.py}); this is a lower bound on the looseness, since an agent may make more calls than a goal requires.

\emph{(iii) The cap.} With fail-closed flagging of episodes that reach $G_{\max}$, $p_{\mathrm{epi}}\le\bar G\,\bar p_{\mathrm{call}}+\Pr(G{>}G_{\max})$; the flag rate is gate-visible, and where padding is used the cap also bounds the per-episode ceiling bill.

\emph{Simulated demonstration} (estimand layer; no measured channel and no real labels, so a construction check only). The setting is $N{=}4000$ episodes, a fraction $0.3$ of them violating with one attempt each, and $k\in\{1,10,100\}$. The coin is a silent per-call coin with $\varepsilon{=}0.2$, and the certifier is the empirical-Bernstein sequence at $\alpha{=}0.05$. The readouts use $g_{\mathrm{assumed}}{=}5$ and $G_{\max}{=}20$, with the level split $\alpha/2$ and a Clopper--Pearson bound on the flag rate. Each row uses $200$ seeds, fixed before running. The static readout collapses as $k$ grows, the measured-$\bar G$ readout stays at about $0.35$, and the capped readout covers by flagging every diluted episode at $k{=}100$ (flag rate $0.30$).

\begin{table}[h]\centering\small
\caption{Dilution: mean readouts of $p_{\mathrm{epi}}{=}0.30$ (coverage).}
\label{tab:dilute}
\setlength{\tabcolsep}{2.5pt}\footnotesize
\begin{tabular}{@{}lccccc@{}}
\toprule
row & true per-call rate & $\bar G$ & static ($g{=}5$) & measured $\bar G$ & capped\\
\midrule
$k{=}1$ & $0.301$ & $1.0$ & $1.000$ (1.00) & $0.350$ (1.00) & $0.358$ (1.00)\\
$k{=}10$ & $0.081$ & $3.7$ & $0.474$ (1.00) & $0.351$ (1.00) & $0.356$ (1.00)\\
$k{=}100$ & $0.010$ & $30.7$ & $0.058$ (0.00) & $0.354$ (1.00) & $0.674$ (1.00)\\
spread & $0.461$ & $1.3$ & $1.000$ (1.00) & $0.669$ (1.00) & $0.674$ (1.00)\\
\bottomrule
\end{tabular}
\end{table}

\section{Additional leakage-assessment details}
\label{app:assess}
\emph{E2 probe-budget sweep.}\label{app:e5sweep} With a noisy clock ($\sigma{=}1$\,s, tool-call), averaging more probes beats the noise. The certificate stays valid through $k\le10^3$, then fails in $77\%$ of seeds at $k{=}10^4$ and in \emph{every} seed at $k{=}10^5$. At that point the true rate has risen to $0.43$ while the certified bound is fooled down to $0.20$. With padding the bound stays at $0.72$, above a true $0.60$, at every $k$ (means over $200$ seeds, $\varepsilon{=}0.2$, $N{=}4000$, normal-mixture sequence). At the channel layer the per-call coin is therefore valid beyond any feasible in-episode probe budget.

\emph{Byte-size dimension.} At fixed $m$, inflating per-object content to $64$\,KB drives one-sided exceedances under the $m$-only ceiling (up to $60/0$ of $60$) and two-sided ones at the largest sizes. A ceiling on total ingested KB $S$, $48.6+5.35\,m+0.144\,S$\,ms, leaves $1$ exceedance in $2{\times}1000$ runs, so the fallback bound of App.~\ref{app:simul} is only $0.15$ at $100$ pairs per point.

\emph{Load.} All ceilings here are constants of one two-core host and must be re-calibrated per deployment. On the six fixed paths at ambient load ($3000$ pairs) there are $0$ exceedances. With $K$ busy cores on the two-core host ($500$ pairs each), $K\le1$ gives $0/500$ in both arms with padded latency independent of $B$ (KS $p{=}0.56$, $0.94$); at saturation ($K{=}2$) the audited arm alone exceeds ($1/500$ under the fixed ceiling; $85$--$99$ per $100$ under the size-aware ceiling, $p_{99.9}$ latency $720$\,ms against a $307$\,ms ceiling at $m{=}96$). In both cases the runtime exceedance count reports the breach ($\DA\le e_0{+}e_1\to1$) and fail-closed counting keeps the certificate sound: load-robustness lives in the counting, not in either ceiling.

\emph{Timestamp clock (E1-ts).} In the two runs of $1000$ pairs the audited inter-timestamp gap has a median of $11.0$\,ms against $0.95$\,ms for the unaudited arm, and $7.1$\,ms against $0.22$\,ms; a held-out classifier reaches $\DA\ge0.99$ in both from one observation (host load average $4.4$ at the end of the run, not controlled).

\emph{CaMeL caveats.} $128$ benign paired checks from $12$ tasks at temperature $0$; checks within a task are correlated ($1$ to $34$ each), so task-level splits give $0.17$ with a cluster-bootstrap interval $[0,0.56]$ that includes zero ($0.19$ $[0,0.63]$ on the $31$ policy checks). Fifteen of $24$ benign runs ended in an interpreter exception, no attacked runs were paired, and there is no padded arm.

\emph{Placement.} A branch moved off the critical path but still concurrent on shared infrastructure re-opens the aggregate-load channel of \S\ref{sec:disc}; the safe placements are strictly post-commit with no agent-readable shared statistic, or pre-commit with the discipline of \S\ref{sec:mechanism}.

\emph{Aggregate load.} Batch-synchronous serving of E1 episodes with $M\in\{2,5,10,20\}$ workers, $300$ rounds per cell, round makespan as the shared statistic, each worker's own coin $B_i\sim\Bern(0.2)$. Single-coin C2ST lower bounds are $0.43$, $0.26$, $0.25$, $0.18$ (CIs from $[0.30,0.56]$ to $[0.14,0.22]$), the audit shifting the statistic by $5.6$--$7.8$\,ms; the measured class moments give $0.48$, $0.28$, $0.19$, $0.13$, below the noiseless projection $0.95$, $0.79$, $0.63$, $0.47$. Capacity isolation (gate workers on one core, audits on the other) halves the shift to $2.7$--$4.3$\,ms but leaves C2ST at $0.33$, $0.33$, $0.17$, $0.13$, since the cores share memory bandwidth and cache. Per-cell load averages ran from $1.6$ to $20$ on the two-core host, so both configurations are oversubscribed at $M\ge10$ and the shared/isolated contrast is indicative only.

\end{document}